\documentclass[preprint,3p]{elsarticle}
\usepackage[pdftex]{hyperref}

\let\today\relax
\makeatletter
\def\ps@pprintTitle{%
	\let\@oddhead\@empty
	\let\@evenhead\@empty
	\def\@oddfoot{\footnotesize\itshape
		{Preprint submitted to CMAME} \hfill\today}%
	\let\@evenfoot\@oddfoot
}
\makeatother

\usepackage[labelformat=simple]{subcaption}
\usepackage{graphics}
\usepackage{epsfig}

\usepackage{float}
\usepackage{amssymb}
\usepackage{amsmath}
\usepackage{amsthm}
\usepackage{amsfonts}
\usepackage{dsfont}
\usepackage{algorithm}
\usepackage{algorithmic}
\usepackage{xcolor}
\usepackage[noabbrev]{cleveref}

\newtheorem{rmrk}{Remark}

\usepackage{mathrsfs}
\usepackage{enumitem}

\graphicspath{{\main/figures/}{figures/}}

\biboptions{sort&compress}

\newcommand{\Vn}{\mathbf{n}}

\newcommand{\R}{\mathbb{R}}

\newcommand{\Sk}{\mathscr{S}}

\newcommand{\V}{\mathbb{V}}

\renewcommand{\div}{\mathrm{div}}

\newcommand{\llbracket}{[\![}
\newcommand{\rrbracket}{]\!]}
\DeclareMathOperator*{\argmin}{arg\,min}

\usepackage{numcompress}

\journal{}

\begin{document}

\begin{frontmatter}
\title{A variational multiscale method derived from an adaptive stabilized conforming finite element method via residual minimization on dual norms}

\author[one,two]{Juan F. Giraldo\corref{mycorrespondingauthor}}
\cortext[mycorrespondingauthor]{Corresponding author}
\ead{jfgiraldoa@gmail.com}
\author[one]{Victor M. Calo}

\address[one]{School of Electrical Engineering, Computing \& Mathematical Sciences, Curtin University, Australia}

\address[two]{Mineral Resources, Commonwealth Scientific and Industrial Research Organisation (CSIRO), Australia}

\begin{abstract}

This paper interprets the stabilized finite element method via residual minimization as a variational multiscale method.
We approximate the solution to the partial differential equations using two discrete spaces that we build on a triangulation of the domain; we denote these spaces as coarse and enriched spaces. 
Building on the adaptive stabilized finite element method via residual minimization, we find a coarse-scale approximation in a continuous space by minimizing the residual on a dual discontinuous Galerkin norm; this process allows us to compute a robust error estimate to construct an on-the-fly adaptive method. We reinterpret the residual projection using the variational multiscale framework to derive a fine-scale approximation. As a result, on each mesh of the adaptive process, we obtain stable coarse- and fine-scale solutions derived from a symmetric saddle-point formulation and an a-posteriori error indicator to guide automatic adaptivity. We test our framework in several challenging scenarios for linear and nonlinear convection-dominated diffusion problems to demonstrate the framework's performance in providing stability in the solution with optimal convergence rates in the asymptotic regime and robust performance in the pre-asymptotic regime. Lastly, we introduce a heuristic  dual-term contribution   in the variational form to improve the full-scale approximation for symmetric formulations (e.g., diffusion problem).

\end{abstract}


\end{frontmatter}

\section{Introduction}
The convection-dominated diffusion equation is a challenging model problem for several physics and engineering applications due to its singularly perturbed behaviour for high P\'eclet numbers, leading to singularities such as sharp inner and boundary layers. Brooks and Hughes~\cite{ Brooks:1982} introduced the streamline-upwind Petrov-Galerkin (SUPG) method to overcome some of these difficulties. This method adds a residual-based stabilizing term to induce a numerical diffusion along the streamlines, which enhances control and stability in the convective operator while conserving consistency in the formulation. The Galerkin/least-squares method (GLS)~\cite{ Hughes:1989} generalizes this idea by adding a least-squares term to the stabilization to enhance control of the whole residual. Although these methods have proven effective at stabilizing the numerical solution of convection-dominated problems, the accuracy of these approaches highly depends on a user-defined stabilization parameter, which requires tuning in most real-world challenges~\cite{John2011}.
In 1998, Hughes~\cite{ Hughes:1998} unified these ideas introducing the variational multiscale method (VMS). This method captures the variational subscales and improves stability properties while maintaining the consistency of the former residual-based methods~\cite{ Hughes:2000, Bazilevs:2007, Bazilevs:2010, Chang:2012, Motlagh:2013, Garikipati:2000, Masud:2009, Masud:2006, Baiges:2017, Codina2000OnSF, Badia2009UnifiedSF, Codina:2018}. VMS decomposes the solution into two scales and keeps their coupling to approximate the fine-scale solution effect that a given mesh cannot capture~\cite{ Hughes2007VariationalMA, Codina2000StabilizationOI, Codina2008AnalysisOA, Codina2002StabilizedFE}. The VMS paradigm led to a reinterpretation of the traditional residual-based methods as approximate sub-grid scale models  and identified the unresolved fine scales as an essential key in the stabilization, even for linear systems including the advection-diffusion equations~\cite{ Hughes:1995, Masud:2004, Hauke2007COMBININGAS}. Related scale-separation models were generalized and extended to diverse applications such as turbulence modeling and wave propagation~\cite{Tran2017, Oberai1998, Sondak2015, Coloms2015AssessmentOV, Akkerman2011, Wang2010}.
\citet{ Cohen:2012} extended the stabilization ideas with least-squares/minimum residual minimization in GLS for more general dual norms. They introduced a saddle point formulation of this residual method, which refers to a VMS formulation. These ideas are extended in the Discontinuous Petrov-Galerkin method (DPG)~\cite{ Demkowicz:2010, Demkowicz:2011, Demkowicz:2011a, Demkowicz2013, Demkowicz2013APD, Carstensen2014, Carstensen2015}  using different non-standard dual norms for stabilization~\cite{ Zitelli:2011, Niemi:2013, Broersen2015, Niemi2011, Niemi2013b, Dahmen2011, Calo2014}. DPG is introduced in the context of a VMS for the convection-diffusion equation in~\cite{ Chan:2013}, offering an alternative, well-behaved sub-grid model to approximate the fine scale ~\cite{ Binev2004, Broersen2015, Dahmen2011}.
The Discontinuous Galerkin Method (dG), initially proposed in 1973 by Reed and Hill~\cite{ Reed:1973}, has offered an alternative stabilization technique over the last few decades and was rapidly extended in numerous applications due to its good stability properties~\cite{ Arnold:2001, Ayuso:2009, Cockburn:2012, Ern:2011, Ern2004TheoryAP}. dG is advantageous over classical methods in providing robustness and high-order accurate approximations, especially for advective operators associated with hyperbolic equations~\cite{ Johnson:1986}. Its stability is induced by the numerical fluxes imposed on internal element interfaces, leading to discretely stable solutions with local conservation. dG is related to residual-based stabilization methods in~\cite{ Johnson:1986, Ern2004TheoryAP, Brezzi:2006} and is used for new stabilized methods based on VMS, including the Multiscale Discontinuous Galerkin method (MSDG) ~\cite{ Bochev:2005, Buffa:2006, Hughes:2006}, the Discontinuous residual-free bubble method~\cite{ Sangalli:2004}, among other recent approaches~\cite{ Coley:2018, Stoter:2022}. 
Alternative stabilization techniques that generalize dG ideas are the Interior Penalty methods that use continuous functions and can handle difficulties encountered by continuous finite element methods in advection-diffusion problems~\cite{ babushka1973InteriorPenalty, douglas1976InteriorPenalty}. These methods penalize flux jumps at mesh interfaces and were applied to biharmonic operators and second-order elliptic and parabolic problems~\cite{ douglas1976InteriorPenalty}. Burman and collaborators have advanced the error estimates for interior penalty methods~\cite{ Burman:2004, Burman:2006} and generalized the ideas to introduce the high-order Continuous Interior Penalty (CIP) finite element method~\cite{ burmanErn2007hpCIP, burmanErn2005hpCIP}. Effectively,~\citet{ burman2009errorCIP} introduced a formulation relating stabilized continuous and discontinuous Galerkin frameworks with the CIP formulation.
Calo et al.~\cite{ Calo:2020} introduced a conforming finite element method that constructs a discrete approximation by minimizing the residual on a dual discontinuous Galerkin norm. This method combines the residual minimization ideas used in DPG with the reliability and inf-sup stability offered by dG. This framework builds on a long tradition of adaptive finite element techniques~\cite{ Demkowicz2002, Vardapetyan1999, Paszyski2006, Binev2002, Binev2004} and were applied to several steady and unsteady partial differential equations~\cite{ Cier:2020, Cier:2021, Calo:2020, Calo:2021, Rojas:2021, Los:2021, Kyburg:2022, Poulet:2022, Giraldo:2023}. We recently used CIP as the coercive discretization for continuous spaces as the underlying stable method to build an adaptive stablized finite element method via residual minimization that uses continuous solutions that require much fewer degrees of freedom than dG on a given mesh~\cite{ Labanda:2022, hasbani2023adaptive}. 
This paper presents a new method integrating ideas from~\cite{ Calo:2020}  to derive a residual minimization approach from a VMS perspective. Our method builds on the concepts from MSDG to decompose a full-discrete discontinuous finite element space into discontinuous (fine) and continuous (coarse) components and leverages VMS analysis to define an inter-scale operator. We begin with an arbitrary discontinuous finite element space and derive a continuous representation by minimizing the residual on a dual discontinuous Galerkin norm. Using the VMS analysis and the residual representative obtained from the dual Galerkin orthogonality, we derive an inter-scale problem to define the stable fine-scale contribution and an error estimator to guide adaptivity. Our approach results in a stable coarse- and fine-scale solution derived from a symmetric saddle-point formulation and an a-posteriori error indicator. We validate our method by tackling challenging linear advection-dominated problems and extend our approach for scalar nonlinear conservation laws by using the Lax-Friedrichs flux within the dG context~\cite{ Juanes:2005, Ern:2011, Vovelle:2002}.
The aims of this paper are: first, to introduce the residual minimization method in~\cite{ Calo:2020}  as a variational multiscale method; secondly, to demonstrate the performance of the fine-scale approximation and its influence on the full-scale approximation in terms of stability and convergence for challenging steady linear and nonlinear convection-dominated diffusion equations. The remainder of the paper is organized as follows: Section 2 introduces the continuous problems and the discontinuous discretization. Section 3 describes the residual minimization strategy. Section 4 presents the variational multiscale formulation. Section 5 shows numerical examples of linear problems. Section 6 extends this approach to nonlinear systems of conservation laws. Finally, we conclude the document with some closing remarks.
\section{Continuous problem and discontinuous discretization} 
\subsection{Advection-diffusion equation}
Let ${\Omega \subset \R^d}$ with $d=2,3$ being a bounded domain with boundary ${\Gamma :=\partial \Omega}$.  Let ${\kappa \in [L^\infty(\Omega)]^{d,d}}$  represent a positive definite diffusion tensor and $\beta \in [L^\infty(\Omega)]^{d}$ a smooth velocity field. Let $\textbf{n}$ be the outward unit normal vector with respect to $\Gamma$; we define the inflow and outflow boundaries, respectively, by
$$\Gamma^- := \left\lbrace x \in \Gamma \, \, | \, \, \beta \cdot \textbf{n} <0 \right\rbrace , \quad \quad \Gamma^+ := \left\lbrace x \in \Gamma \, \, | \, \, \beta \cdot \textbf{n} \geq 0 \right\rbrace. $$
We denote by $\Gamma_D$ the Dirichlet and $\Gamma_N$ the Neumann boundary functions, which are a complementary subset of $\Gamma$ (i.e.,$\Gamma = \overline{\Gamma_N \cup \Gamma_D} $). 
Thus, we define the inner and outer parts of the Neumann 
 boundary as follow:
\begin{equation}\label{eq:Outer-inner}
	\begin{aligned}
    	\Gamma_N^- &:=&  \Gamma_N \cap \Gamma^- , \quad \quad \Gamma_N^+ := \Gamma_N \cap \Gamma^+.
	\end{aligned}
\end{equation}
We consider the following advection-diffusion equation in strong form as:
\begin{equation}\label{eq:DAR}
    \begin{aligned}
    -\div \left( \kappa \nabla u \right) + \beta \cdot \nabla u &= f&& \text{ in } \Omega, \smallskip\\
    u &= u_D && \text{ on } \Gamma_D,\\
    (-\beta u+\kappa \nabla u) \cdot \textbf{n}    &=h_N&&  \text{ on } \Gamma^-_N, \\
    \kappa \nabla u \cdot \textbf{n}    &=h_N&& \text{ on } \Gamma^+_N, \\
\end{aligned}
\end{equation}
where we denote $f\in L_2(\Omega)$ as the source term and $u_D\in H^{-1/2}(\Gamma_D)$  and $h_N\in H^{-1/2}(\Gamma_N)$ as the Dirichlet and Neumann boundary functions, respectively. Following the function spaces $V:= \left\lbrace v\in H^1(\Omega): v|_{\Gamma_D}=0 \right\rbrace$ and $\Tilde{V}:= \left\lbrace u\in H^1(\Omega): u|_{\Gamma_D}=u_D \right\rbrace$,  the continuous weak variational formulation of~\eqref{eq:DAR} reads:
\begin{equation}\label{eq:scalar}
	\left\{
	\begin{array}{l}
		\text{Find } u \in \Tilde{V} \text{ such that:} \smallskip \\
		\begin{aligned}
			(\nabla v,\kappa \nabla u)_{\Omega} + (v,\beta \cdot \nabla u)_{\Omega} +  (v,(\beta\cdot\textbf{n})u)_{\Gamma^-} = (v,f)_{ \Omega} + 
			(v,h_N)_{\Gamma_N}  , \quad \forall v \in V,
		\end{aligned}
	\end{array}
	\right.
\end{equation}
where $(\cdot,\cdot)_{\Omega}$ and $(\cdot,\cdot)_{\Gamma}$ denote the $L_2$-scalar product on $\Omega$ and ${\Gamma}$, respectively.

\subsection{Discontinuous Galerkin (dG) discretization}
Let $\mathfrak{T}$ be a coarse triangulation, such that $\Omega$ is decomposed into \textit{n} subdomains $K$ as  $\mathfrak{T}:\{K_i\}^n_{i=1}$. We define the broken space of discontinuous functions as: 
\begin{equation}
\label{eq:Vh_space}
\V_h:= \left\lbrace v \in L_2(\mathfrak{T}): v | _{K} \in \mathbb{P}^p(K) \in ,\forall K \in  \mathfrak{T} \right\rbrace,
\end{equation}
where $\mathbb{P}^p$ denotes the set of functions of degree lower than or equal to $p$.
Let $K_1$ and $K_2$ represent two disjoint elements in $\mathfrak{T}$, and let $F$ be their shared face. We define the set of all faces as $\Sk_h:= \bigcup_{K \in \mathfrak{T}} F$. We further define the set of interior faces by $\Sk^0_h=\Sk_h \backslash \Gamma$ and the set of boundary faces by $\Sk^\partial_h = \Sk_h \cap \Gamma$. We also define the set of inflow boundary faces as $\Sk_h^{\partial^-}:= \Sk_h^\partial \cap\Gamma^-$, the set of outflow boundary faces as $\Sk_h^{\partial^+}:= \Sk_h^\partial \cap\Gamma^+$, and we denote by $\Sk^D_h=\Sk_h \cap \Gamma^D$ and $\Sk^N_h=\Sk_h \cap \Gamma^N$ the sets of Dirichlet and Neumann faces, respectively.
Let $h_K$ be the element diameter, $h_F$ the face diameter, and $\textbf{n}_F$ be the unit normal vector in the outer direction of $F$ from $K_1$ to $K_2$. Given a scalar field $v$ and denoting $v_{1,2}:=v|_{K_{1},K_{2}}$, we define the arithmetic average $\{v\}$, weighted average $\{v\}_{\omega}$ and jump $[\![v]\!]$ on an internal face $F\in\Sk^0_h$ by
$$ \{v\}:= \frac{1}{2}(v_1 + v_2),      \quad  \quad \{v\}_{\omega}:=v_1\omega_1+ v_2\omega_2,      \quad  \quad  [\![v]\!] := v_1 - v_2,$$
where $\omega_1+\omega_2=1$ and $\omega_1,\omega_2 \geq 0$. 
For heterogeneous diffusion, we use:
\begin{equation}
\begin{aligned}\label{eq:omega1}
\omega_1 = \frac{\delta_1}{\delta_{1} + \delta_{2}}, \quad \quad \omega_2 = \frac{\delta_2}{\delta_{1} + \delta_{2}},
\end{aligned}
\end{equation}
with $\delta_{1}=  \textbf{n}_F  \cdot \kappa_{1} \textbf{n}_F $ and   $\delta_{2}=\textbf{n}_F \cdot \kappa_{2} \textbf{n}_F $. In the homogeneous diffusion case, these weights reduce to $\omega_1 = \omega_2 = \frac{1}{2}$, recovering the arithmetic average. We set on boundary faces $F\in\Sk^{\partial}_h$ that $\{v\}=\{v\}_{\omega}=[\![v]\!]=v$. For further details in the discrete setting, see~\cite{ Ern:2009}.
We use the following dG discrete form:
\begin{equation}
\label{eq:DG_prob}
\left\{
\begin{array}{l}
\text{Find } u_h  \in  \V_h  , \text{such that:}   \smallskip \\
\begin{aligned}
 b_h(v_h,u_h) = \ell_h(v_h) \quad \forall \, v_h \in \V_h,
\end{aligned}
\end{array}
\right.
\end{equation}
where $b_h$ defines the dG discrete bilinear form considering the contribution of the diffusive part from the Symmetric Weighted Interior Penalty form (SWIP) and the advective part from the upwinding formulation:
\begin{equation}
    b_h(v,u) := b_h(v,u)^{\text{swip}} + b_h(v,u)^{\text{upw}},
\label{eq:a_h}
\end{equation}
with, 
\begin{align}
    b_h(v,u)^{\text{swip}}&:=    \sum_{K \in \mathfrak{T}} (\nabla_h v , \kappa \, \nabla_h u )_{K} \nonumber\\
    &- \sum_{F \in \Sk_h^0}\Big(([\![ v ]\!]\, , \,\{ \kappa \nabla_h u\}_{\omega}\cdot \Vn_F)_F + ( \{ \kappa\nabla_h v\}_{\omega}\cdot \Vn_F \, , \, [\![ u ]\!] )_F - \eta_e \gamma_{\kappa}([\![ v ]\!],[\![ u ]\!])_F \Big)\nonumber\\
	&- \sum_{F \in \Sk_h^D}\Big(( v  , \,\kappa \nabla_h u\cdot \Vn_F)_F + ( \kappa\nabla_h v\cdot \Vn_F \, , \,  u  )_F - \eta_e \gamma_{\kappa}(v,u )_F\Big),
	\label{eq:swip}\\
	b_h(v,u)^{\text{upw}}&:= \sum_{K \in \mathfrak{T}} \left( v , \beta \cdot \nabla u \right) _K + \sum_{F \in \Sk_h^{\partial^{-}}} 
	\left( v , \left(\beta \cdot \textbf{n}_F \right) u \right)_F \nonumber\\  
 &+ \sum_{F \in \Sk_h^0} \Big(  \left( [\![ v ]\!] ,\tfrac{1}{2} |\beta \cdot \textbf{n}_F|  [\![ u ]\!] \right) _F  -\left ( \{ v \} ,  (\beta \cdot \textbf{n}_F)  [\![ u ]\!] \right) _F \Big),
	\label{upw}
\end{align}
where $(\cdot ,\cdot)_F$ and  $(\cdot,\cdot)_K$ represent the inner product over the discrete face and internal element, respectively. The diffusion penalty $\gamma_{\kappa}$ is defined for all internal faces $F\in\Sk^0_h$ as: $$\gamma_{\kappa}:= \dfrac{2\kappa_1\kappa_2}{\kappa_1+\kappa_2},$$ 
and $\eta_e$ denotes a positive penalty defined as~\cite{ Shahbazi:2005}:
\begin{equation}
    \eta_e := \frac{(p+1)(p+d)}{d}	\left\{
    \begin{aligned}
    &\tfrac{1}{2} \left (\tfrac{\mathcal{A}(\partial
	   K_1)}{\mathcal{V}(K_1)}
            + \tfrac{\mathcal{A}(\partial K_2)}{\mathcal{V}(K_2)}\right ),
    &&\text{if} \quad F = \partial K_1 
        \cap \partial K_2,\\
    &\tfrac{\mathcal{A}(\partial K)}{\mathcal{V}(K)},
    &&\text{if} \quad F = \partial K
    \cap \Gamma    , 
    \end{aligned}\right.
\label{Eta_constant}
\end{equation}
where $d$ denotes the problem dimension, and $p$ represents the polynomial degree of the test space. In 3D, $\mathcal{V}$ and $\mathcal{A}$ denote the volume and area of an element, respectively. In 2D, they represent the length and area of the element. The right-hand side in~\eqref{eq:DG_prob} reads for weakly imposed non-homogeneous Dirichlet and Neumann boundary conditions as:
\begin{align}
	\ell_h(v):
	= \sum_{K \in \mathfrak{T}}(v,f)
	&+ \sum_{F \in \Sk_h^{D}} \left( \eta_e\kappa(v, u_D)_F -  ( \kappa\nabla_h v \cdot \Vn_F , u_D)_F \right) \nonumber\\ 
	&-\sum_{F \in \Sk_h^{D} \cap\Gamma^-} \left(v ,\left( \beta \cdot
	\textbf{n}_F \right) u_D \right) _F + \sum_{F \in \Sk_h^N} (v , h_N) _F.
	\label{lineear_form}
\end{align}
We endow $\V_h$ with the norm:
\begin{align}\label{eq:normVh}
\|w\|^2_{\V_h} := \|w\|^2_{\text{upw}} + \|w\|^2_{\text{swip}},
\end{align}
with 
\begin{align}
\|w\|^2_{\text{adv}} &:= \displaystyle \|w\|^2_{0} + \displaystyle \sum_{ F \in \mathscr{S}_h}\left( \tfrac{1}{2} | \, \beta \cdot \Vn_F | \, \llbracket w \rrbracket, \llbracket w \rrbracket \right) _{0,F} + \displaystyle \sum_{K \in \mathfrak{T}} h_K \| \, \beta \cdot\nabla{w} \, \|^2_{0,K},
\label{adv-norm}\\
\|w\|^2_{\text{swip}} &:= \displaystyle \| \, \kappa \nabla w \, \|^2_{0} +  \sum_{F \in \mathscr{S}_h} \left( \eta_e \kappa \llbracket w \rrbracket, \llbracket w \rrbracket \right) _{0,F},
\label{eq:swip-norm}
\end{align}
where $\|\cdot\|_{0}$ represents the $L_2$-norm on the domain $\Omega$ and $\|\cdot\|_{0,\Gamma}$ on its boundary $\Gamma$, while $\|\cdot\|_{0,K}$ and $\|\cdot\|_{0,F}$ denote the $L_2$-norm on the element $K$ and  faces $F$, respectively.  
Problem~\eqref{eq:DG_prob} is well-posed, and the inf-sup stability is established through the norm~\eqref{eq:normVh}. Refer to~\cite{ Ayuso:2009} for further details.  
The sub-index h for the test and trial solutions (i.e., $u_h$ and $v_h$) are dropped in the following sections to keep the notation simple.

\section{Residual minimization formulation}
\label{Section:Residual-minimization}
This section briefly recaps the residual minimization strategy described in~\cite{ Calo:2020}. The main idea is to deliver a stable approximation in a continuous space by minimizing the residual onto a discontinuous Galerkin norm. First, we define a subspace  $\bar \V_h \subset \V_h$, such that $\bar \V_h$ is the $H^1$-conforming subspace. Then, from the stable formulation in~\eqref{eq:DG_prob}, we use the trial subspace $\bar \V_h$ to solve the following minimization problem:
 \begin{equation}
 	\label{eq:min_problem}
 		\left\{
 	\begin{array}{l}
	\text{Find } \bar u  \in \bar \V_h \subset \V_h , \text{such that:}   \smallskip \\
	\begin{aligned}
		\bar u = \displaystyle \argmin_{z \in \bar \V_h } \dfrac{1}{2}\|\ell_h- B_h \, z\|^2_{\V_h^\ast}, 
	\end{aligned}
\end{array}
	\right.
 \end{equation}
where the operator $B_h$ is defined by $\langle \cdot , B_h w \rangle_{\V_h \times \V_h^*} := b_h(\cdot,w), \forall w \in \V_h.$ 
We state~\eqref{eq:min_problem} as a critical point of the minimizing functional, which can be translated into the following linear problem:
 \begin{equation}
	\label{eq:min_problem2}
	\left\{
	\begin{array}{l}
		\text{Find } \bar u  \in \bar \V_h , \text{such that:}   \smallskip \\
		\begin{aligned}
( B_h\delta u  ,\ell_h-B_h \bar u)_{\V_h^*} =g(R^{-1}_{\V_h} B_h\delta u  ,R^{-1}_{\V_h}(\ell_h-B_h \bar u)) = 0  \quad \forall \delta u \in \bar \V_h,
		\end{aligned}
	\end{array}
	\right.
\end{equation}
where $g(\cdot, \cdot)$ represents the inner product that induces the discrete norm ${\|\cdot\|_{\V_h}}$ and $R_{\V_h}$ denotes the Riesz operator which maps the elements in $\V_h$ to the dual space $\V_h^\ast$, such that: 
\begin{equation}\label{eq:rieszmap}
  \begin{aligned}
  \langle \cdot , R_{\V_h} y  \rangle_{\V_h \times \V_h^\ast}:= g(\cdot,y).
  \end{aligned}
\end{equation}
Defining the residual representation function as
$$\varepsilon := R^{-1}_{\V_h}(\ell_h -B_h \bar u)\in \V_h,$$
which implies
$$g(\varepsilon,v)=l_h(v) -b_h(u,v), \quad \forall v \in \V_h;$$
thus, the residual minimization problem in~\eqref{eq:min_problem} leads to the saddle point problem:
\begin{equation}
	\label{eq:saddle-point1}
	\left\{
	\begin{array}{l}
		\text{Find } (\varepsilon, \bar u) \in \V_h \times \bar \V_h,  \text{ such that:} \smallskip \\
		\begin{array}{rcll}
			g(v,\varepsilon) + b_h(v,\bar u) + b_h(\varepsilon,\bar w) &=& l_h(v),& \quad \forall (v,\bar w) \in  (\V_h,\bar \V_h), \smallskip\\
		\end{array}
	\end{array}
	\right.
\end{equation}
where orthogonality condition between the residual representation and the $H^1-$conforming subspace of $\V_h$ is enforced as $b_h(\varepsilon,\bar w)=0$. Features of the methodology and the properties of the error estimator and continuous solution are discussed in detail in~\cite{ Calo:2020, Cier:2021} in the context of advection-reaction-diffusion equations. 

\section{A variational multiscale interpretation of the residual minimization framework}

\subsection{A multiscale partition of the trial and test spaces} 

Following the variational multiscale arguments, we decompose $\V_h$ into coarse and fine scales. Since we use a Petrov-Galerkin formalism, our test and trial spaces are different. Thus, starting from the $H^1-$conforming solution space $\bar \V_h$ and define appropriate direct sum partitions of the entire function space $\V_h$ using the operators from the residual minimization framework. 

First, we define the space $\V_h '$ as the annihilator of the bilinear form $b_h(\cdot,\bar w) \ \forall \bar w \in \bar\V_h$ (e.g., the residual representative belongs to this set of linear functionals that map the operator's range to zero). Thus, given $\bar \V_h \subset \V_h$, we define $\V_h'$ as the annihilator of the bilinear form $b_h$ acting on $\V_h$, such that
\begin{align}\label{eq:fine-scale}
	\V_h' :=\{v \in \V_h \ | \ b_h(v,\bar w)=0,\qquad  \forall \, \bar w \in \bar \V_h \},
\end{align}
Next, we define $\hat \V_h$ as the orthogonal complement of $\V_h'$ with respect to the inner product $g(\cdot,\cdot)$, that is:
\begin{align}\label{eq:coarse-scale}
    \hat \V_h :=\{v \in \V_h \ | \ g(v,v')=0,\qquad  \forall \, v' \in \V_h' \}.
\end{align}
Last, we define the complement of the coarse-scale solution space as the kernel of the bilinear $b_h$ when tested by $\hat\V_h$. Thus, given $\hat \V_h \subset \V_h$, we define $\tilde \V_h$ to be
\begin{equation}\label{eq:fine-scale2}
	\begin{array}{l}
		\tilde \V_h :=\{u \in \V_h \ | \ b_h(\hat v,u)=0,\qquad  \forall \, \hat v \in \hat \V_h \}.
	\end{array}
\end{equation}
where the (full-scale) trial space is $\V_h := \bar \V_h  \oplus \tilde \V_h$ and the (full-scale) test space is $\V_h := \hat \V_h  \oplus \V'_h$. 

We have a direct sum decomposition of the trial space to deliver a solution $u \in \V_h$ in~\eqref{eq:DG_prob}, such that:
\begin{equation}\label{eq:Direct-sum}
u = \bar u + \tilde u.
\end{equation}
where we denote $\bar u \in \bar \V_h$ and $\tilde u \in \tilde\V_h$ as the coarse- and  fine-scale trial functions. Similarly, for any test function $v \in \V_h$, we can write 
\begin{equation}
    v = \hat v + v',
    \label{eq:v_space}
\end{equation}
where we denote $\hat v \in \hat \V_h$ and $v' \in \V_h '$ as the coarse- and  fine-scale test functions. Thus, we have two complementary direct sum decompositions of $\V_h$. 

Thus, the saddle point problem in~\eqref{eq:saddle-point1} can be stated as follows:
\begin{equation}
\label{eq:res-minimization2}
	\left\{
	\begin{array}{l}
		\text{Find } (\varepsilon', \bar u)  \in (\V_h' \times \bar \V_h ), \text{such that:}   \smallskip \\
		\begin{aligned}
			g(v,\varepsilon') + b_h(v,\bar u)  = \ell(v), &&  \forall v \in \V_h 
		\end{aligned}
	\end{array}
	\right.
\end{equation}
Splitting the test function using~\eqref{eq:v_space}, we reformulate problem~\eqref{eq:res-minimization2} as two complementary but independent problems for the coarse-scale solution and the fine-scale residual representative:
\begin{itemize}
\item Residual reconstruction (fine-scale problem) as
    \begin{equation}\label{eq:Error-finescale}
	\left\{
	\begin{array}{l}
		\text{Find } \varepsilon'  \in \V_h' , \text{such that}   \smallskip \\
		\begin{aligned}
			g(v',\varepsilon')  = \ell(v'), \quad \quad  \forall v' \in \V_h' ,&& 
		\end{aligned}
	\end{array}
	\right.
\end{equation}
\item $H^1-$conforming solution (coarse-scale problem) as
\begin{equation}\label{eq:Coarse-scale}
	\left\{
	\begin{array}{l}
		\text{Find } \bar u  \in \bar U, \text{such that}   \smallskip \\
		\begin{aligned}
			b_h(\hat v,\bar u)  = \ell(\hat v), \quad \quad  \forall \hat v \in \hat \V_h, && 
		\end{aligned}
	\end{array}
	\right.
\end{equation}
which corresponds to a Petrov–Galerkin method with optimal test functions~\cite{ Demkowicz:2011, Zitelli:2011, Niemi:2013, Calo2014}.
\end{itemize}

From the definitions of the complementary direct sum decompositions~\eqref{eq:fine-scale}-\eqref{eq:fine-scale2}, we have that  $b_h(\hat v ,\tilde u ) = b_h(v',\bar u)  = 0 $; thus, problem~\eqref{eq:DG_prob} can be split into the following two problems:
\begin{equation}\label{eq:vf_cont_split}
	\left\{
	\begin{array}{l}
		\text{Find } (\bar u,\tilde u) \in (\bar \V_h, \tilde \V_h), \text{ such that:} \smallskip \\
        \begin{aligned}
    		b_h(\hat v,u) &:= b_h( \hat v ,\bar u) =\ell(\hat v ), && \qquad\forall \hat v \in \hat \V_h,  \\
	   	b_h(v',u) &:=  b_h(v',\tilde u) = \ell(v'),  &&\qquad\forall v' \in \V_h'. 
        \end{aligned}
	\end{array}
	\right.	
\end{equation}
Consequently, from~\eqref{eq:Error-finescale} and~\eqref{eq:vf_cont_split}$_2$, we can rewrite the fine-scale component of the discrete solution in terms of the residual error estimate
\begin{align}\label{eq:vf_cont_split2}
b_h(v',\tilde u) =  g(v',\varepsilon')= \ell(v'), \qquad \forall v' \in \V_h'.  
\end{align}
Thus, we can express~\eqref{eq:vf_cont_split}$_2$ such that the fine-scale solution satisfies the following problem:
\begin{equation}\label{eq:res-dgprime1}
	\left\{
	\begin{array}{l}
		\text{Find } \tilde u  \in \tilde\V_h , \text{  such that:  }   \smallskip \\
		\begin{aligned}
			b_h(v',\tilde u)  = g(v',\varepsilon'),&& \forall v' \in \V_h'.
		\end{aligned}
	\end{array}
	\right.
\end{equation}
Lastly, since $b(\hat v,\tilde u ) = g(\hat v,\varepsilon')  = 0$,  we can express the fine-scale problem equivalently as
\begin{equation}\label{eq:res-gral1}
	\left\{
	\begin{array}{l}
		\text{Find } \tilde u  \in \V_h , \text{  such that:  }   \smallskip \\
		\begin{aligned}
			b_h(v,\tilde u)  = g(v,\varepsilon'), && \forall v \in \V_h.  
		\end{aligned}
	\end{array}
	\right.
\end{equation}
We solve the discrete problem resulting from the system~\eqref{eq:res-gral1} rather than~\eqref{eq:res-dgprime1} or~\eqref{eq:vf_cont_split}$_2$ since the discrete solution is identical, simpler to implement, and cheaper to compute.

\begin{rmrk}[Relation between VMS reconstruction \& dG solution]
By construction, the partitioned full-scale approximation in~\eqref{eq:Direct-sum} is identical to the classical dG solution.  Next, we introduce an adjoint multiscale reconstruction; thus, we denote by ($\theta$) the dG solution of~\eqref{eq:DG_prob}.
\end{rmrk}

\subsection{Adjoint multiscale reconstruction} 

Using the direct sum partitions of the test and trial spaces, we exploit the insight behind the adjoint residual-based estimator proposed for goal-oriented adaptivity in~\cite{ Rojas:2021}; therein; the authors obtained a residual representative for the quantity of interest by solving a well-posed ad hoc discrete problem. In the present context, this adjoint residual problem is driven by $\varepsilon'$. We introduce the adjoint reconstruction by revisiting the multiscale partitions from the previous section. Given $\varepsilon'$ that solves~\eqref{eq:Error-finescale}, this error representation $\varepsilon'$ is proportional to the discrete system's residual~\cite{ Calo:2020, Rojas:2021}. From the definition of $\V_h'$, we know that
$$b_h(\varepsilon',\bar v)=0,\qquad\qquad\forall\bar v\in\bar\V_h.$$
Also, from~\eqref{eq:res-gral1}, we have that
$$b_h(v,\tilde u)  = g(v,\varepsilon'), \qquad \qquad  \forall v \in \V_h;$$
next we add a heuristic interaction of $\varepsilon'$ with the whole test space $\V_h$ to the fine-scale driving force. Thus, we postulate the following problem, where the heuristic fine scales $\check u\in\V_h'$ solve
\begin{equation}\label{eq:res-dgprime}
	b_h(v,\check u)  = g(v,\varepsilon')+b_h(\varepsilon',v) \quad \quad  \forall v \in \V_h.
\end{equation}
Thus, this heuristic fine-scale postprocessing of the error representative includes two extra contributions:
\begin{align}\label{eq:res-dgprime-fine}
    b_h(\hat v,\check u) &= b_h(\varepsilon',\hat v) &&\forall \hat v \in \hat \V_h, \\
    b_h(v',\check u)  &= \ell_h(v') + b_h(\varepsilon',v') && \forall v' \in \V_h',
\end{align}
where the first equation contributes to the coarse-scale trial space in $\bar\V_h$ while the second one contributes to the fine-scale trial space $\V_h'$.

In short, we propose the following heuristic adjoint variational multiscale reconstruction such that:
$$\phi = \bar u + \check u,$$
where $\check u \in \V_h$ is the reconstructed fine-scale solution that solves~\eqref{eq:res-dgprime} for a given error estimate $\varepsilon'$. 
In the next section, we test the performance of $\phi$ in the $L_2$ and energy $(\V_h)$ norms, showing an improvement in the asymptotic regime, especially for diffusion-dominated problems (i.e., $||u_{exa} - \phi ||_{\V_h} \lesssim ||u_{exa} - u ||_{\V_h} $).

\section{Numerical examples}\label{sec:examples}

\begin{figure}[h!]
	\centering
	\begin{tabular}{ c c c }
		\hspace{-0.2cm}
		\includegraphics[scale=0.06] {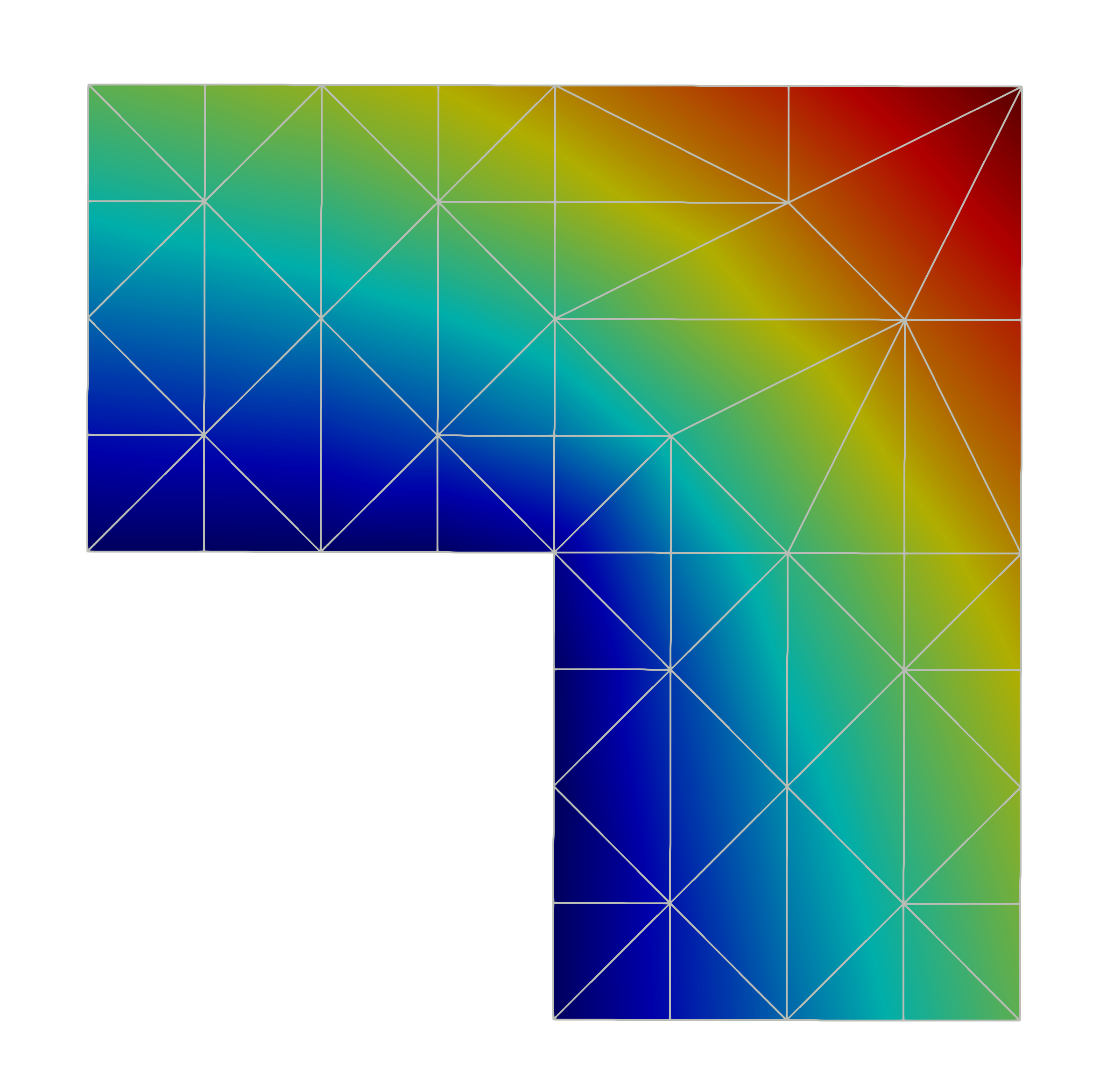}  &
		\includegraphics[scale=0.06] {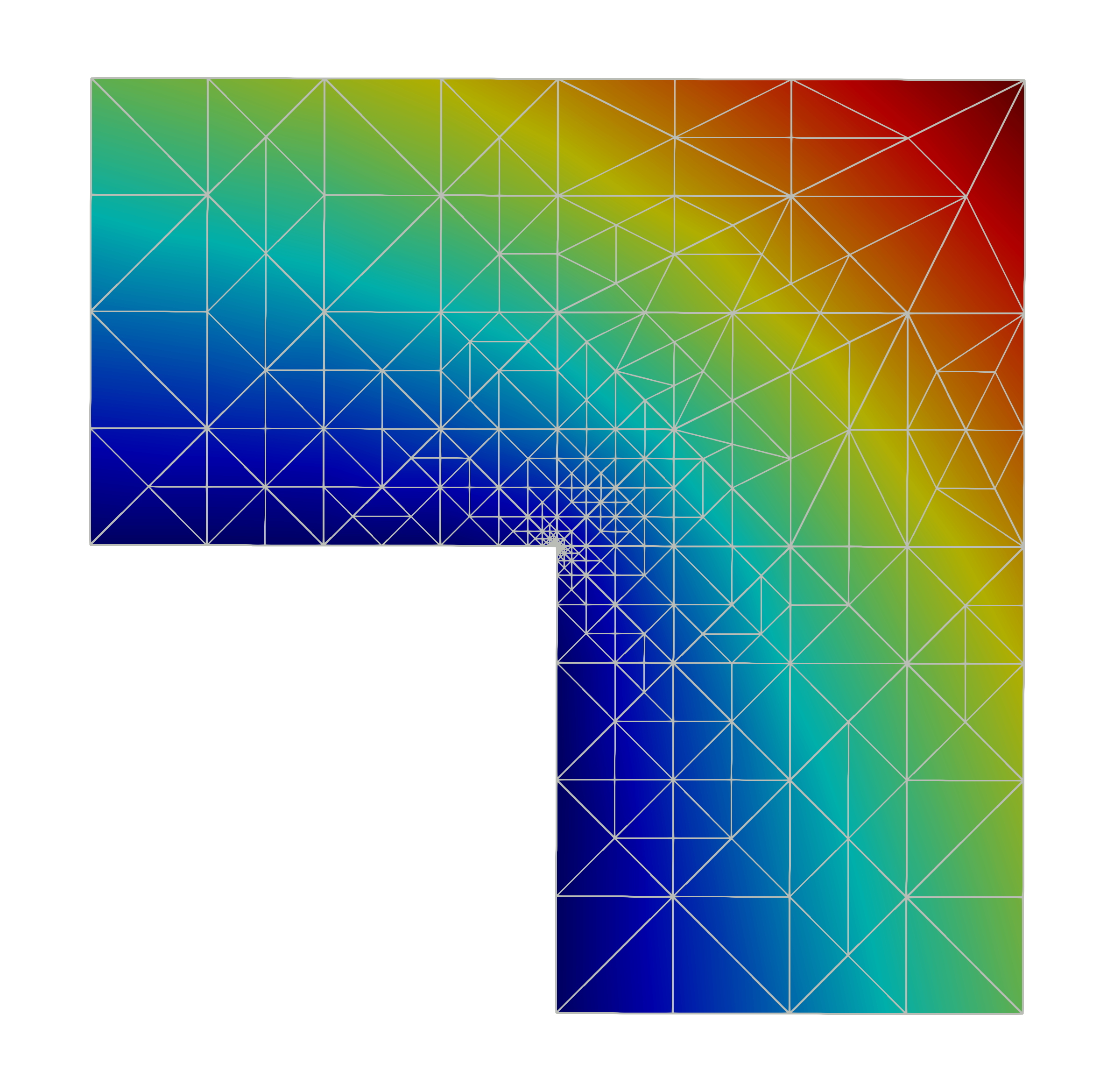}  &
		\includegraphics[scale=0.06] {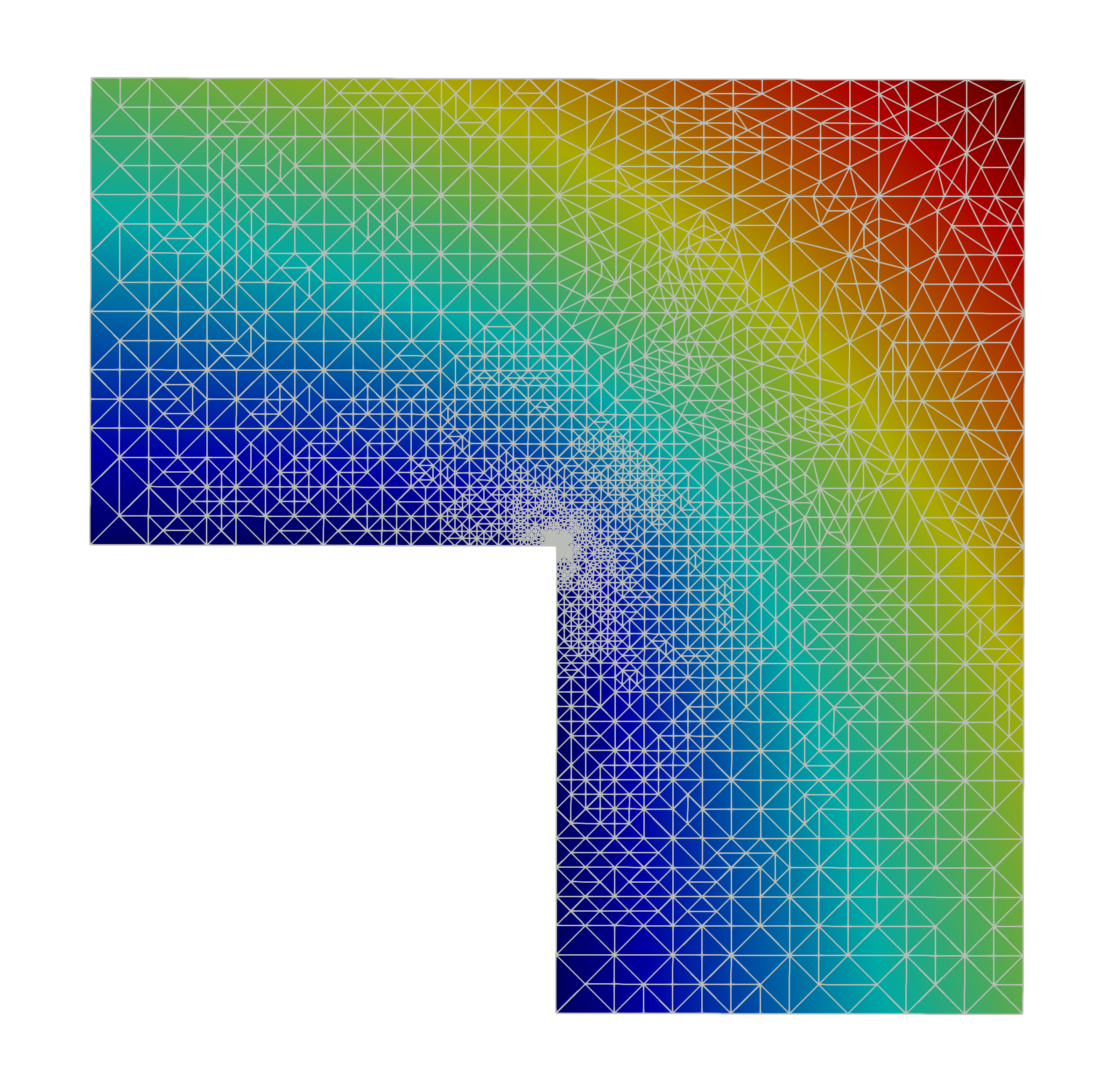} \\
		\footnotesize {Level= 0} & \footnotesize {Level = 10} & \footnotesize {Level = 20}
	\end{tabular}
	\caption{Solution for the re-entrant corner problem for different refinement levels}
	\label{L-shape-solutions}
\end{figure}

\subsection{Refinement strategy}
Next, we demonstrate the efficacy of our methods in different linear cases through a range of numerical examples. We analyze the rate decay for both the coarse and reconstructed full-scale solutions in the $L_2$ and energy ($\V_h$) norms and compare these results with the classical dG solution.  The convergence plots in the following sections show the error norm versus the total number of degrees of freedom ($DoF^{1/d}$) (i.e., $\dim(\bar{\V}_h)+\dim(\V_h)$ ).
We validate our formulation by evaluating and comparing the performance of our multiscale approach using some test problems described in~\cite{ Cier:2021}.

We implement an adaptive refinement marking strategy for the following examples using an extended version of the D\"orfler bulk-chasing criterion.
Our adaptive procedure considers an iterative process in which each iteration consists of the following four steps:
\begin{equation}
    \textrm{SOLVE} \rightarrow \textrm{ESTIMATE} \rightarrow \textrm{MARK} \rightarrow \textrm{REFINE}
\end{equation}
We mark the elements' contribution to a user-defined fraction ($\eta_{ref}$) of the total estimated error $||\varepsilon||^2_{\V_h}$.  Let $\eta_{ref}$ be $0.25$ and $0.125$ for 2D and 3D problems, respectively (see~\cite{ Giraldo:2023} for implementation details).
We solve the saddle point problem in~\eqref{eq:saddle-point1} employing an iterative algorithm described in~\cite{ Bank:1989, Calo:2020} and use FEniCS~\cite{ Alnaes:2015} as a platform to perform all the numerical simulations. 


\begin{figure}[h!]
	\begin{subfigure}{.49\textwidth}
		\centering
		\includegraphics[width=1.04\textwidth]{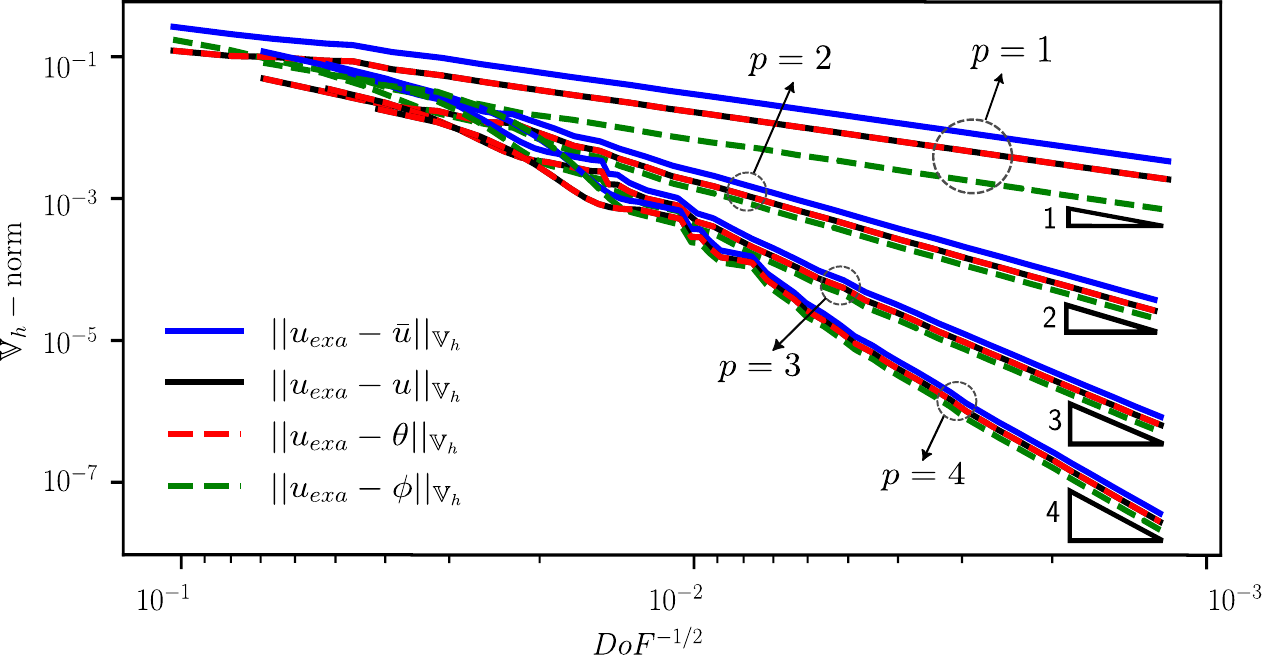}
		\caption{$\V_h$ norm}
		\label{fg:L-shape_Vh}
	\end{subfigure}
	\begin{subfigure}{.49\textwidth}
		\centering
		\includegraphics[width=1.04\textwidth]{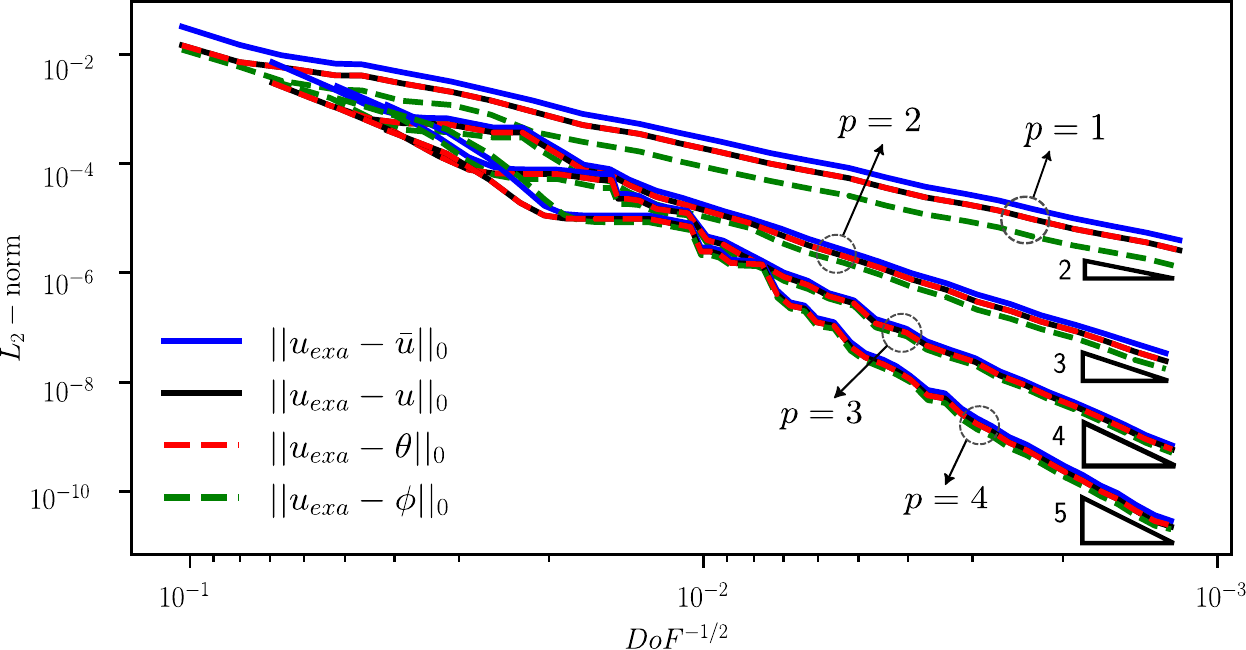}		
		\caption{$L_2$ norm}
		\label{fg:L-shape_L2}
	\end{subfigure}
    \caption{$\V_h-$ \& $L_2-$norm convergence for Laplace problem}
	\label{fg:L-shape-convergence}
\end{figure}

\subsection{Diffusion problem for a domain with a re-entrant corner}\label{ss:Lshape}

We study the method's performance with the diffusion problem in a re-entrant corner domain~\cite{ Mitchell:2013}. The problem configuration induces a singularity at the inward-pointing vertex of the concave polygon. Since capturing corner singularities is challenging for uniform refinement techniques, this problem tests the adaptive grid refinement algorithms. We solve the following Laplace equation in an L-shape domain $\Omega$:
\begin{align*}
    \Delta {u} &= 0  , &&\text{in } \Omega= (-1,1)^2 \backslash (-1,0]^2, \\
    u &= u_D  , &&\text{on } \partial\Omega=\Gamma_D,
\end{align*} 
where the Dirichlet boundary conditions ($u_D$) are applied based on the exact solution:
$$u_{exa} =  r^{\alpha}\sin({\alpha \theta}) , $$
with $r = \sqrt{x^2+y^2}$, $\theta = \tanh^{-1}(y/x)$ and $\alpha = 2/3$.\\
Figure~\ref{L-shape-solutions} shows three surface plots of the full-scale solution and the corresponding meshes for different adaptive refinement levels.  The results show the error estimator's robustness and the energy norm's effectiveness at capturing the singularity. Convergence plots are presented in Figure~\ref{fg:L-shape-convergence} for different test-function polynomial degrees ($p=1,2,3,4$) in the $L_2$ and energy norms. Here, we show optimal rate decay for coarse $\bar{u}$ and full-scale $\tilde{u}$ solutions and verify that the fine-scale contribution $\tilde{u}$ recovers the dG approximation $\theta$ (i.e., $u \approx \theta$ ). Besides, Figure~\ref{fg:L-shape-convergence} shows that the adjoint multiscale reconstruction ($\phi$) improves the approximation compared to the full-scale $(u)$ and the dG $(\theta)$ solutions regardless of the polynomial degree.  

\begin{figure}[h!]
	\centering
	\begin{subfigure}[b]{0.48\textwidth}  
		\centering 
    	\includegraphics[width=1.04\textwidth]{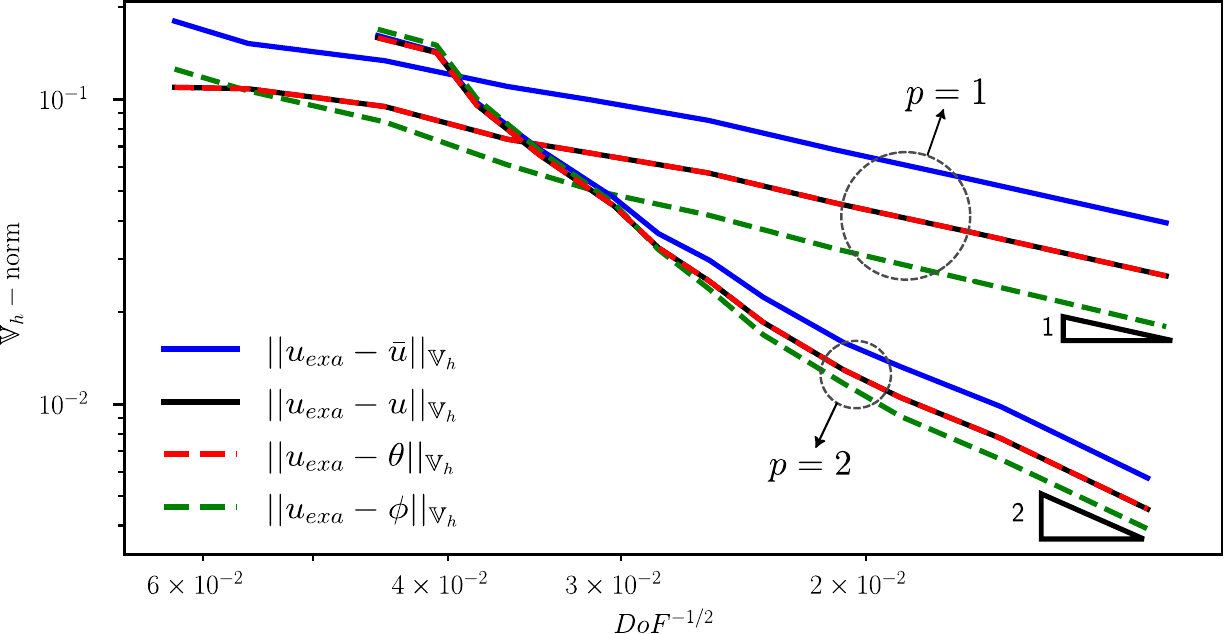}
		\caption{$\V_h$ norm}
		\label{fg:Fichera-Vh-q10}
	\end{subfigure}
    \hfill
	\begin{subfigure}[b]{0.48\textwidth}
        \centering		
        \includegraphics[width=1.04\textwidth]{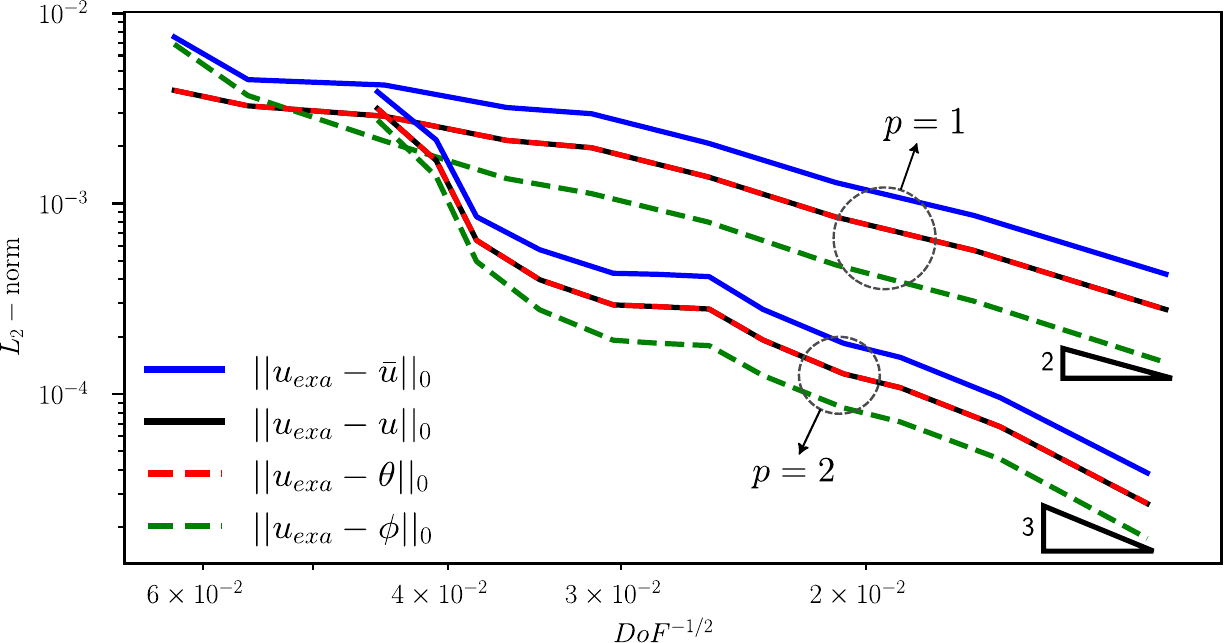} 
		\caption{$L_2$ norm}
		\label{fg:Fichera-L2-q10}
	\end{subfigure}
	\caption{$\V_h-$ and $L_2-$norm convergence for Fichera corner, $q=\dfrac{1}{10}$}
	\label{fg:Fichera-problem-q10}
\end{figure}

\begin{figure}[h!]
	\begin{subfigure}[b]{0.48\textwidth}   
		\centering 		\includegraphics[width=1.04\textwidth]{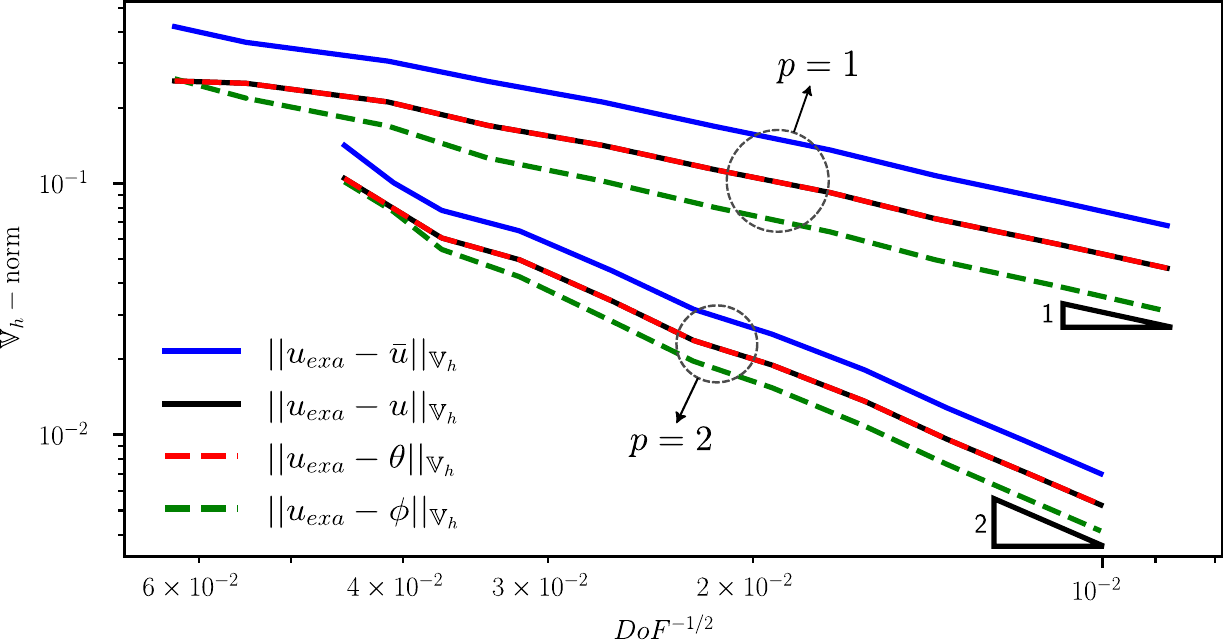}
		\label{fg:Fichera-Vh-q3}
	\end{subfigure}
    \hfill
	\begin{subfigure}[b]{0.48\textwidth} 
		\centering 		
        \includegraphics[width=1.04\textwidth]{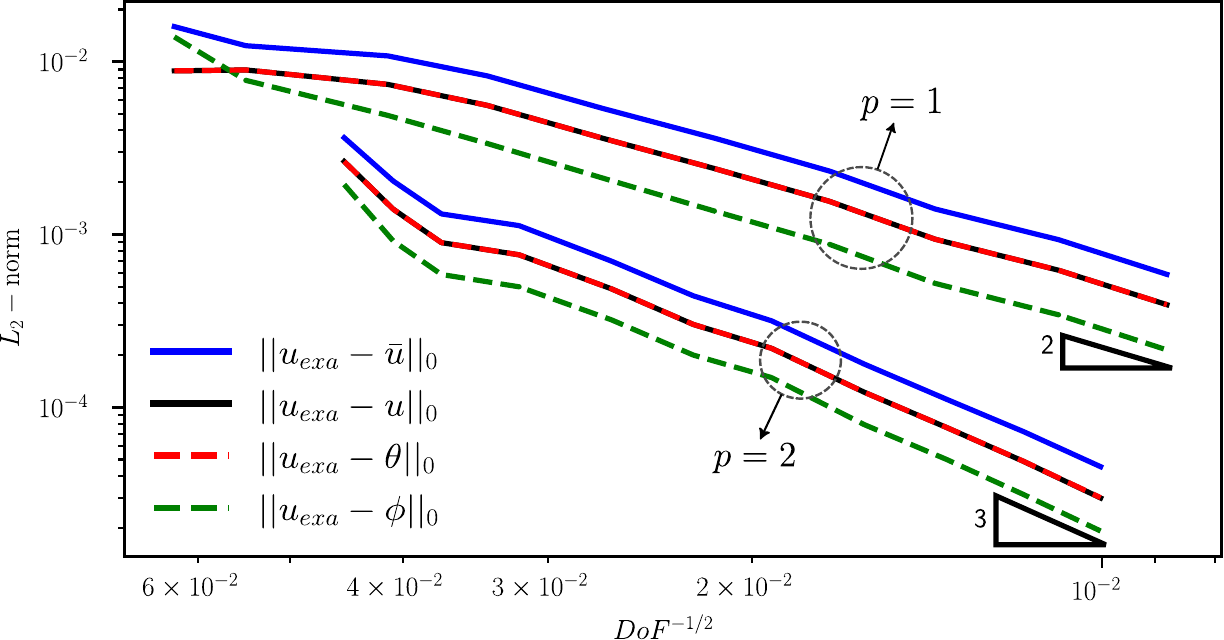}
		\caption{$L_2$ norm}
		\label{fg:Fichera-L2-q3}
	\end{subfigure}
	\caption{$\V_h-$ and $L_2-$norm convergence for Fichera corner, $q=\dfrac{1}{3}$}
	\label{fg:Fichera-problem-q3}
\end{figure}

\subsection{Diffusion problem for a 3D domain with a Fichera corner}

In this example, we extend results in Section~\ref{ss:Lshape} for the 3D Fichera corner problem, where we induce the singularity at the re-entrant corner of the domain  $\Omega = (-1,1)^3 \backslash [0,1)^3$.  We consider the problem:
\begin{equation}
\begin{array}{rlll}
-\kappa \Delta {u} &=& f  , \quad &\text{in } \Omega, \\
u &=& u_D  , \quad &\text{on } \Gamma_D,
\end{array}
\label{eq:poisson}
\end{equation}  
where $\kappa=1$ and source term $f$ and Dirichlet boundary conditions $(u_D)$ are derived from the exact solution:
 $$u_{exa} = \left (\sqrt{x^2+y^2+z^2}\right)^q$$
with $q=1/10$ and $q=1/3$.
Similarly to the 2D case, uniform refinement techniques may struggle to accurately capture the behavior in the Fichera corner. The high spatial gradients and the necessity for fine meshes at the singularity lead to suboptimal convergence when no adaptive refinement techniques are used~\cite{ Calo:2020}. Figures~\ref{fg:Fichera-problem-q10} and~\ref{fg:Fichera-problem-q3} display the convergence plots in $L_2$ and $\V_h$ norms. Here, we test the robustness of the error estimator to provide optimal convergence rates for the coarse solution and to recover the dG solution and optimality in the full-scale solution for different $q$ values. Moreover, as in Section~\ref{ss:Lshape}, the ajoint multiscale reconstruction ($\phi$) provides a better approximation compared to the full-scale solution ($u$) in the $L_2$ and $\V_h$ norms, and improves the pre-asymptotic convergence rates, especially for linear trial functions.

\begin{figure}[h!]
	\begin{center}
		\begin{subfigure}{0.49\textwidth}
			\centering
			\includegraphics[width=1.01\textwidth]{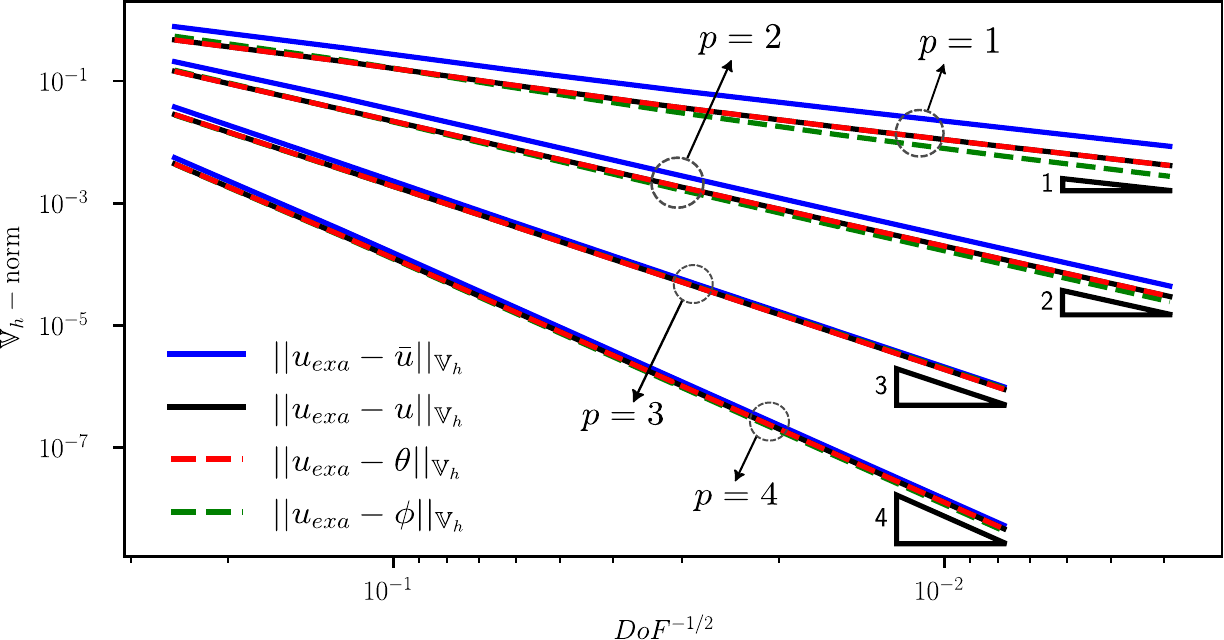}
			\caption{$\V_h$ norm}
			\label{fg:HeterogenousVh}
		\end{subfigure}
		\begin{subfigure}{0.49\textwidth}
			\centering
			\includegraphics[width=1.01\textwidth]{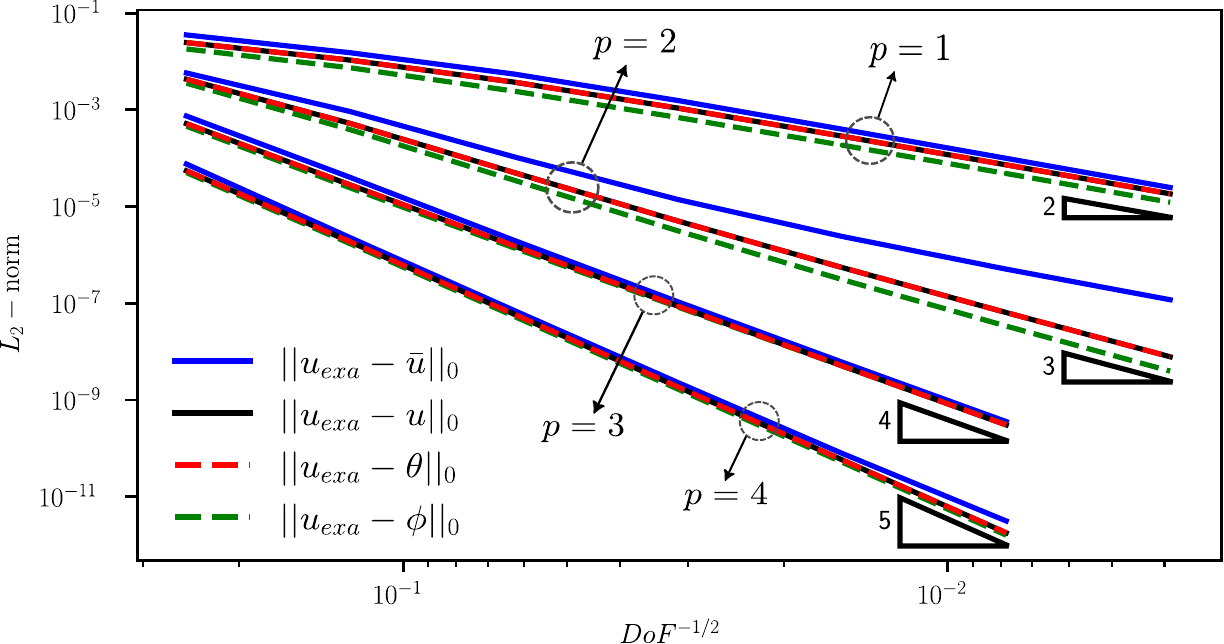}	
			\caption{$L_2$ norm}
			\label{fg:HeterogenousL2}
		\end{subfigure}
	\end{center}
	\caption{$\V_h-$ \& $L_2-$norm convergence for heterogeneous diffusion problem}
    \label{fg:heterogeneous-problem}
\end{figure}

\subsection{Heterogeneous Diffusion problem}\label{ss:hetero2D}

We solve the advection-diffusion equation with heterogeneous and anisotropic diffusion, following~\cite{ Cier:2021}:
\begin{equation}\label{eq:DAR-HD}
    \begin{aligned}
        -\kappa{\Delta u} + \beta \cdot \nabla u &= 0,&& \text{ in } \Omega, \smallskip\\
        u &= u_D, && \text{ on } \Gamma,\\
    \end{aligned}
\end{equation}
where $\beta=(0,1)^T$ and $\kappa$  is a second rank tensor with different $\epsilon_i$ values for each domain: 
\begin{equation*}
    \kappa|_{\Omega_i}=
    \begin{pmatrix}
        \epsilon_i & 0 \\
        0 & 1.0
    \end{pmatrix}.
\end{equation*}
We partition $\Omega=[0,1]^2$ into two domains:  $\Omega_1=[0,\frac{1}{2}]\times [0,1]$ and $\Omega_2=[\frac{1}{2},1]\times[0,1]$, such that $\epsilon_1=0.1$ and $\epsilon_2=1.0.$
 We impose Dirichlet boundary conditions ($u_D$) based on the exact solution for each domain~\cite{ Burman:2006}:
 \begin{equation}
    u_{exa} = \left\{\begin{array}{rl}  \left(u_{1/2}-\exp\left(\frac{1}{2\epsilon_1}\right)  + (1-u_{1/2})\exp\left(\frac{x}{\epsilon_1}\right)\right)/\left(1-\exp\left(\frac{1}{2\epsilon_1}\right) \right)& \text{if} \quad  x \in \Omega_1,\\
 
  \left(-\exp\left(\frac{1}{2\epsilon_2}\right)u_{1/2} + u_{1/2} \exp\left(\frac{x-\frac{1}{2}}{\epsilon_2}\right)\right)/ \left(1-\exp\left(\frac{1}{2\epsilon_2}\right)\right)  & \text{if} \quad x \in \Omega_2,  \end{array} \right .
 \end{equation}
where 
\begin{align}
    u_{1/2}=\left( \frac{\exp\left(\frac{1}{2\epsilon_1}\right)}{1-\exp\left(\frac{1}{2\epsilon_1}\right)}  \right)\left( \frac{\exp\left(\frac{1}{2\epsilon_1}\right)}{1-\exp\left(\frac{1}{2\epsilon_1}\right)} +\frac{1}{1-\exp\left(\frac{1}{2\epsilon_2}\right)} \right)^{-1}.
\end{align}
Figure~\ref{fg:heterogeneous-problem} plots the convergence in the $L_2$ and $\V_h$ norms for polynomial orders~1 to~4. As noted in~\cite{ Cier:2021}, Figure~\ref{fg:HeterogenousL2} shows a loss in the convergence rate in the $L_2$ norm for the coarse-scale solution ($\bar u$) for even polynomial degrees. However, we can recover dG optimality by including the fine-scale contribution ($u'$), regardless of the polynomial degree. The adjoint multiscale reconstruction ($\phi$) is more accurate than the other discrete approximations on the same mesh.

\begin{figure}[h!]
	\begin{center}
		\begin{subfigure}{0.495\textwidth}
			\centering
			\includegraphics[width=\textwidth]{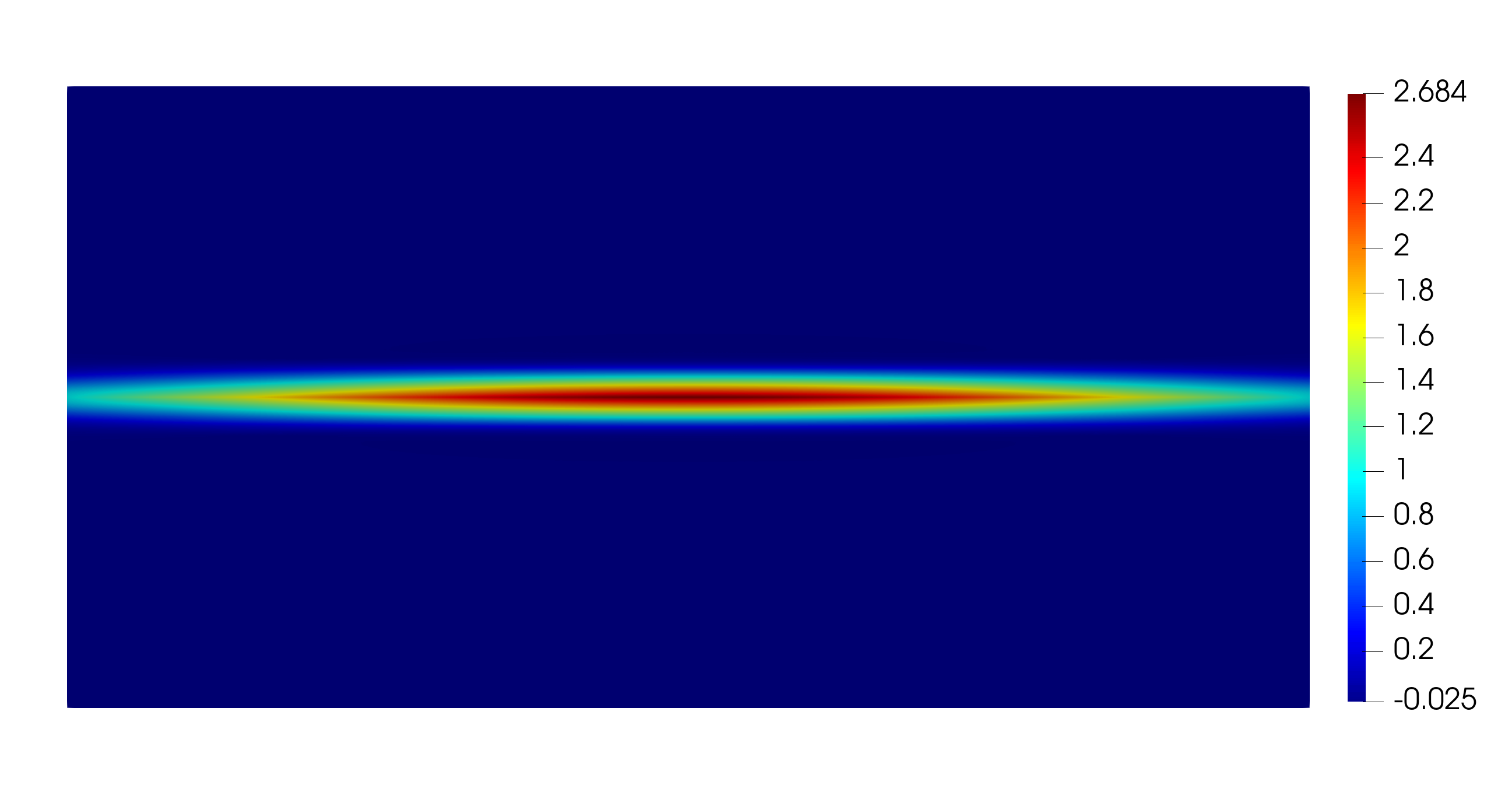}
			\caption{Coarse-scale solution}
			\label{fg:Anisotropic-Coarse}
		\end{subfigure} 
		\begin{subfigure}{0.495\textwidth}
			\centering
			\includegraphics[width=\textwidth]{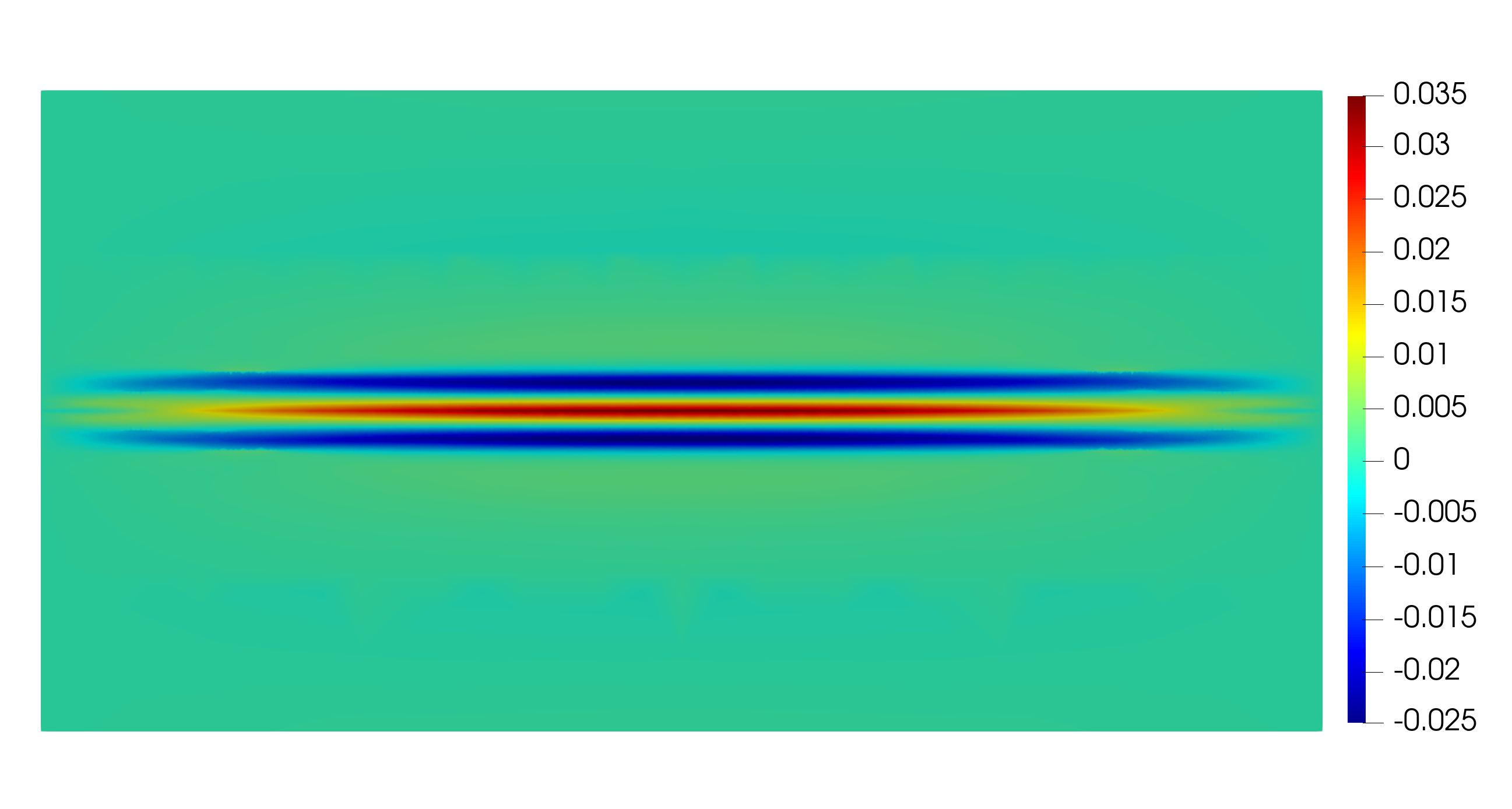}
			\caption{Fine-scale solution}
			\label{fg:Anisotropic-Fine}
        \end{subfigure} 
    	\begin{subfigure}{0.5\textwidth}
    	  \centering
            \hspace{-1cm}
    	   \includegraphics[width=\textwidth]{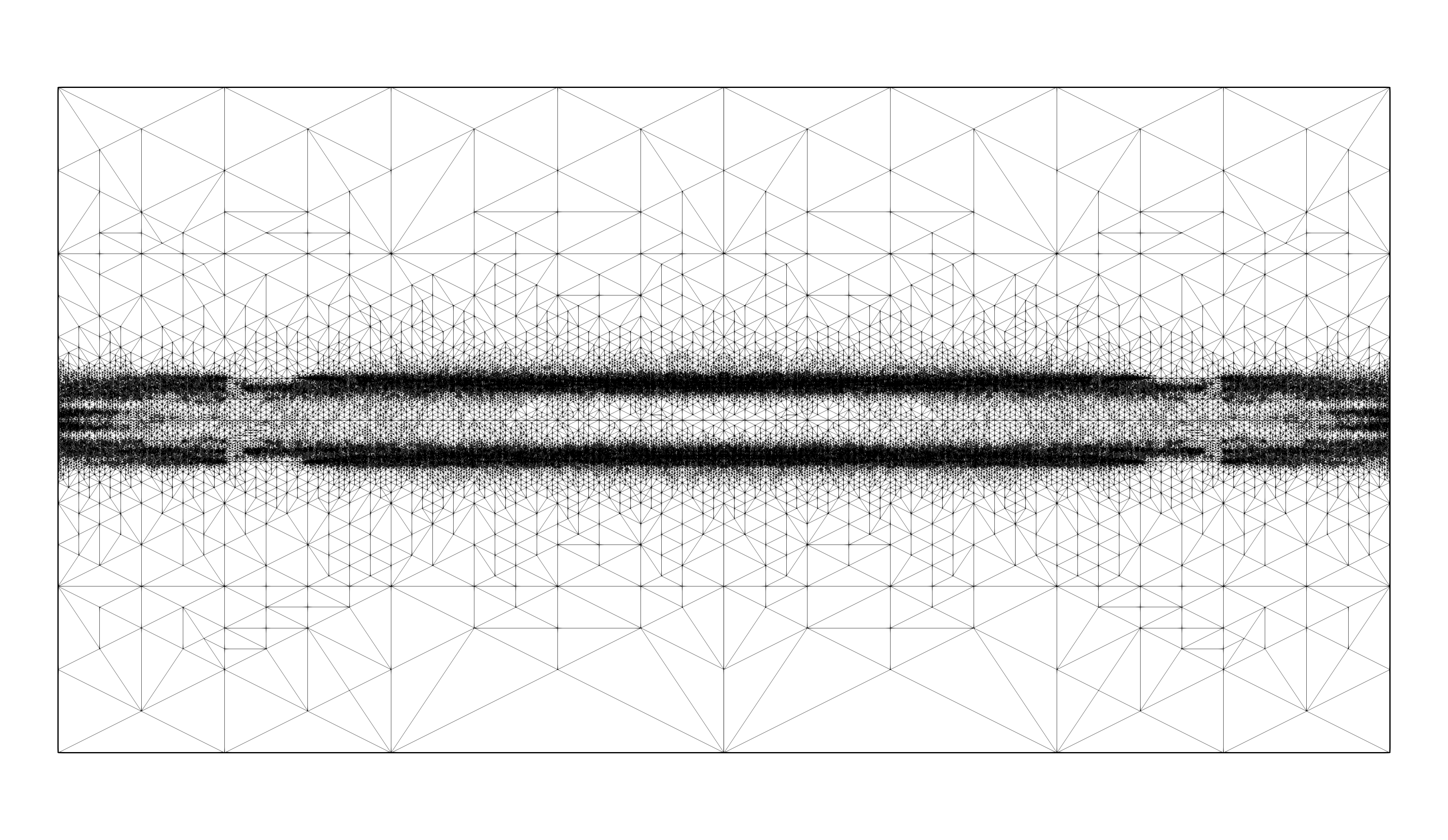}
            \caption{Refined mesh}
       	\label{fg:Anisotropic-Mesh}
        \end{subfigure} 
        \caption{Coarse- \& fine-scale solutions with final mesh for strongly anisotropic diffusion problem with $r=10^6$ \& $p=1$}
        \label{fg:Anisotropic-Solution}
	\end{center}
\end{figure}
%


\subsection{Strongly anisotropic diffusion}
In this example, we test the performance of our method in a highly anisotropic diffusion problem. We solve equation~\eqref{eq:poisson} with high contrast in the permeability tensor:
\begin{equation}
	\kappa|_{\Omega}=\begin{pmatrix}
		\alpha_{\kappa} & 0 \\
		0 & \gamma_{\kappa}
	\end{pmatrix}.
\label{eq:ani-tensor}
\end{equation}
We define the anisotropy ratio, $r_{\kappa}$, as the ratio between the maximum and minimum values of the diffusion coefficients; thus, we set $r_{\kappa}:=\alpha_\kappa/\gamma_\kappa$ for~\eqref{eq:ani-tensor}.  High $r_{\kappa}$ values are challenging in this problem, corresponding to locally small weights in the diffusion tensor, leading to advection-dominated regimes.
We study the method's performance by solving~\eqref{eq:ani-tensor} for different values of the anisotropy ratio by imposing a sharp inner layer problem based on the following Gaussian-function-type manufactured solution, as described in~\cite{ Pestiaux:2014}:
\begin{equation}
u_{exa} = \frac{\exp({-[ x^2 + r_{\kappa}\tau y^2 ]})}{4\pi\sqrt{r_{\kappa}\tau}},
\label{eq:gaussian-solution}
\end{equation}
with $\tau = 10^{-3}$. We use~\eqref{eq:gaussian-solution} and $\alpha_{\kappa} = 1$ to derive the source term ($f$) and Dirichlet boundary conditions ($u_D$) in the domain $\Omega = [-1,1]\times[-0.5,0.5]$. 

\begin{figure}[h!]
	\begin{center}
		\begin{subfigure}{0.495\textwidth}
			\centering
			\includegraphics[width=\textwidth]{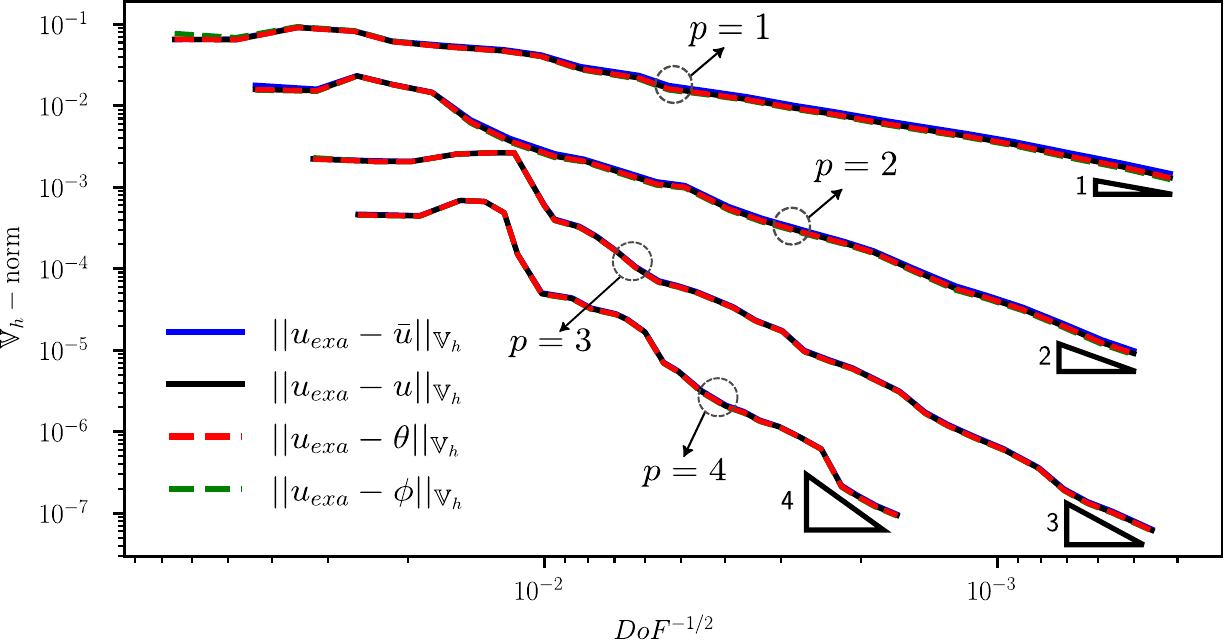}
			\caption{$\V_h$ norm}
			\label{fg:Anisotropic-vms-Vh-1e4}
		\end{subfigure}
		\begin{subfigure}{0.495\textwidth}
			\centering
			\includegraphics[width=\textwidth]{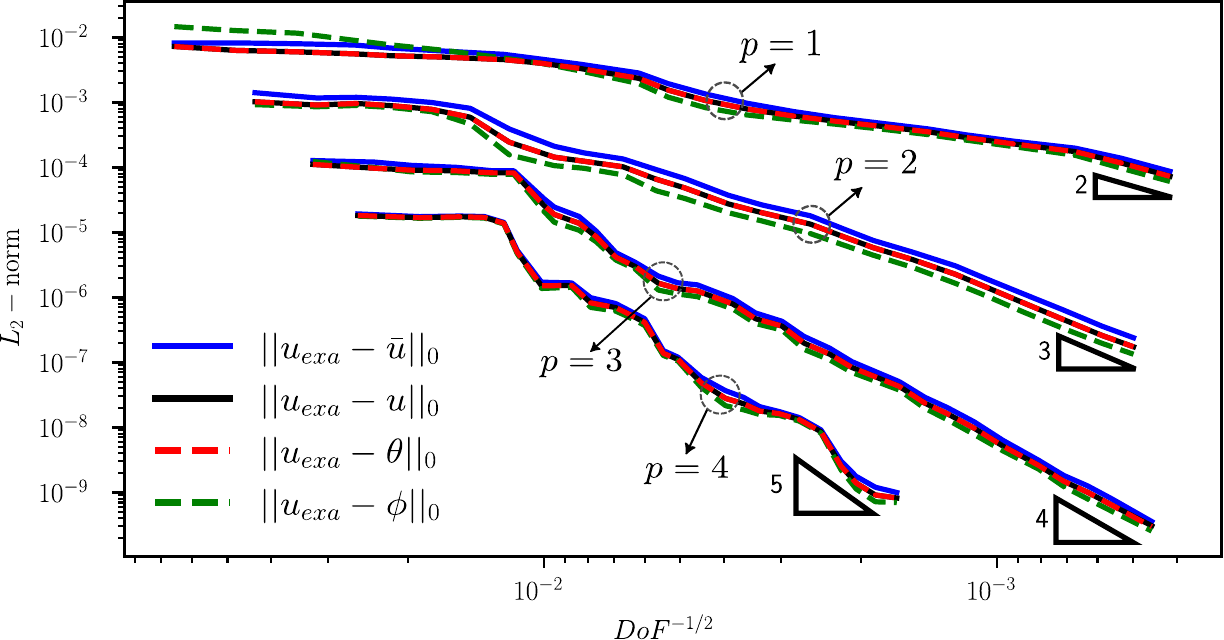}
			\caption{$L_2$ norm}
			\label{fg:Anisotropic-vms-L2-1e4}
		\end{subfigure}
	    \caption{$\V_h-$ and $L_2-$norm convergence for anisotropy ratio $r_{\kappa}=10^4$.}
	    \label{fg:anisotropic-1e4}
	\end{center}
\end{figure}

\begin{figure}[h!]
	\begin{center}
		\begin{subfigure}{0.495\textwidth}
			\centering
			\includegraphics[width=\textwidth]{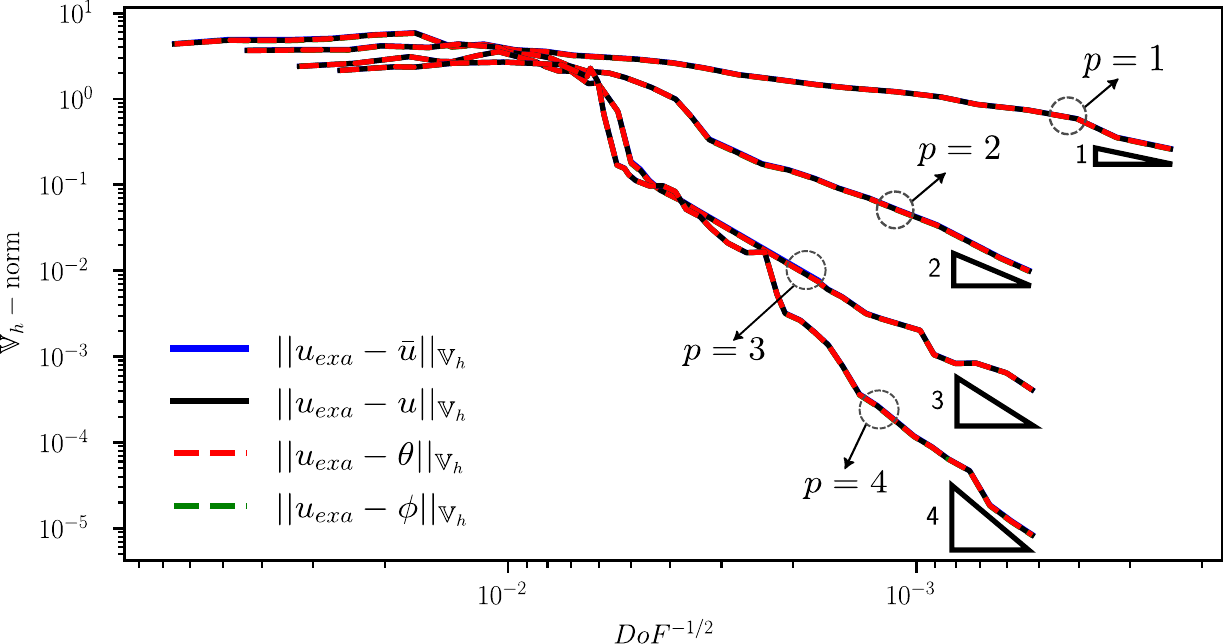}
			\caption{$\V_h$ norm}
			\label{fg:Anisotropic-vms-Vh-1e6}
		\end{subfigure}
		\begin{subfigure}{0.495\textwidth}
			\centering
			\includegraphics[width=\textwidth]{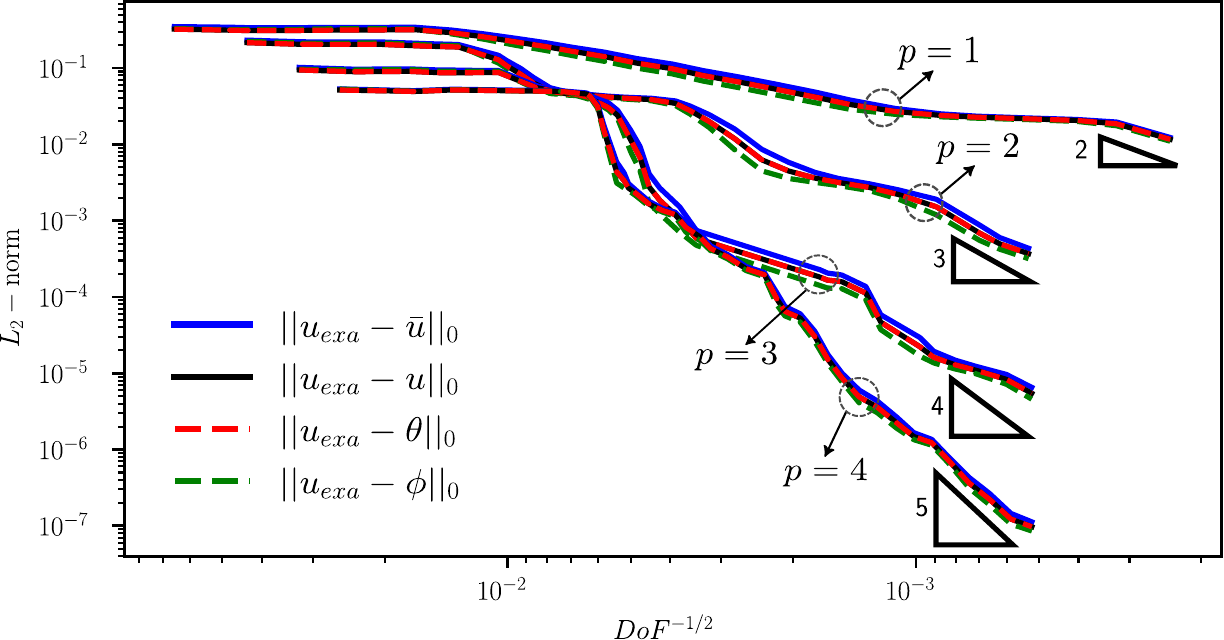}	
			\caption{$L_2$ norm}
			\label{fg:Anisotropic-vms-L2-1e6}
		\end{subfigure}
	    \caption{$\V_h-$ and $L_2-$norm convergence for anisotropy ratio $r_{\kappa}=10^6$. }
		\label{fg:anisotropic-1e6}
	\end{center}
\end{figure}
\noindent Figure~\ref{fg:Anisotropic-Solution} presents the coarse- and fine-scale solutions showing the discontinuity in $y=0$ and the robustness in the error estimator to effectively refine the inner layers. Figures~\ref{fg:anisotropic-1e4} and~\ref{fg:anisotropic-1e6} show the convergence plots for $r_{\kappa} =10^4$ and $r_{\kappa} =10^6$, respectively. We obtain optimal rates for different polynomial degrees in the $\V_h$ and $L_2$ norms for all discrete approximations on the refined-mesh sequences.

\begin{figure}[h!]
	\begin{center}
		\centering
		\includegraphics[width=.875\textwidth]{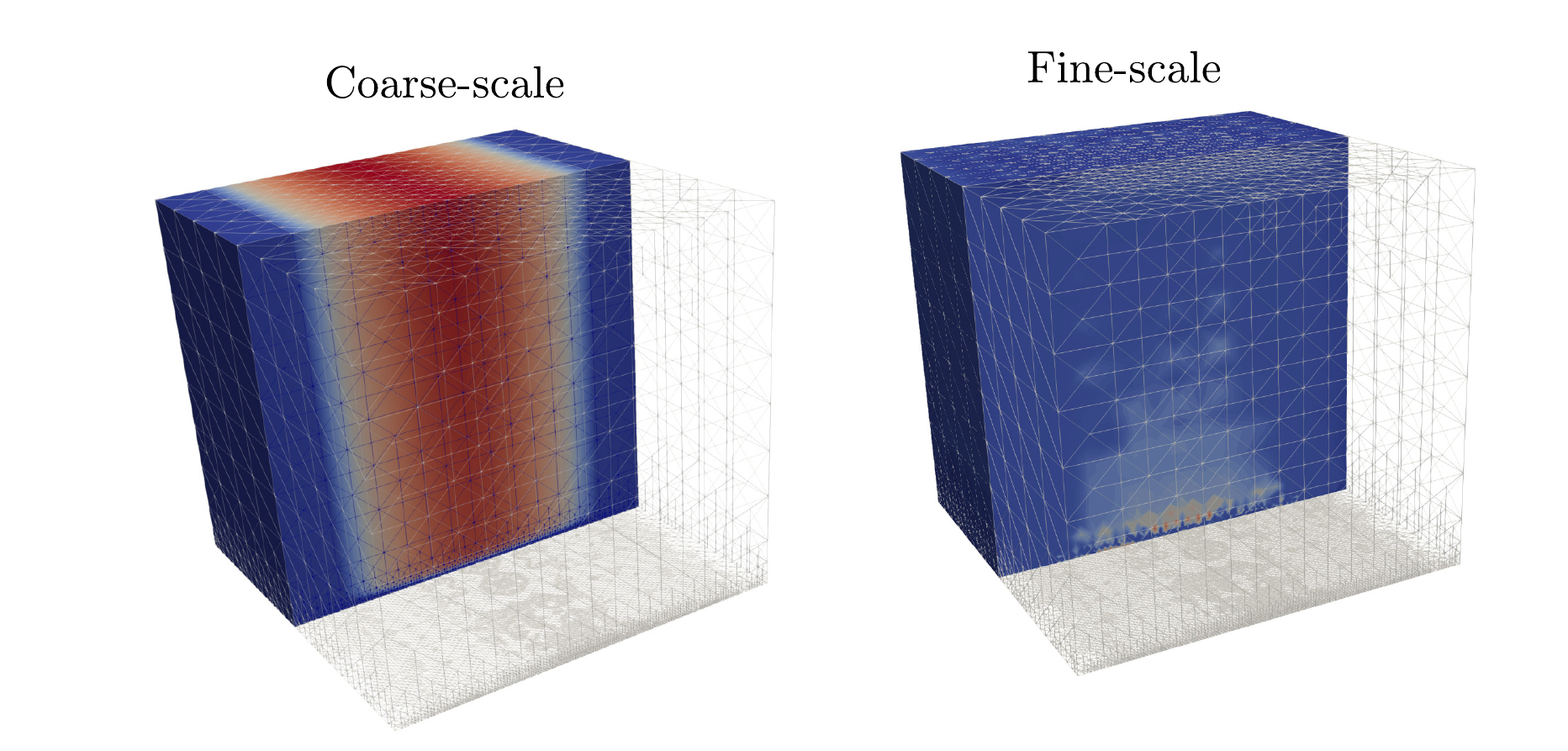} 
		\caption{Coarse- \& fine-scale approximations for the 3D Eriksson-Johnson problem}
		\label{fg:EJD}
	\end{center}
\end{figure}

\begin{figure}[h!]
	\begin{center}
		\begin{subfigure}{0.495\textwidth}
			\centering
			\includegraphics[width=\textwidth]{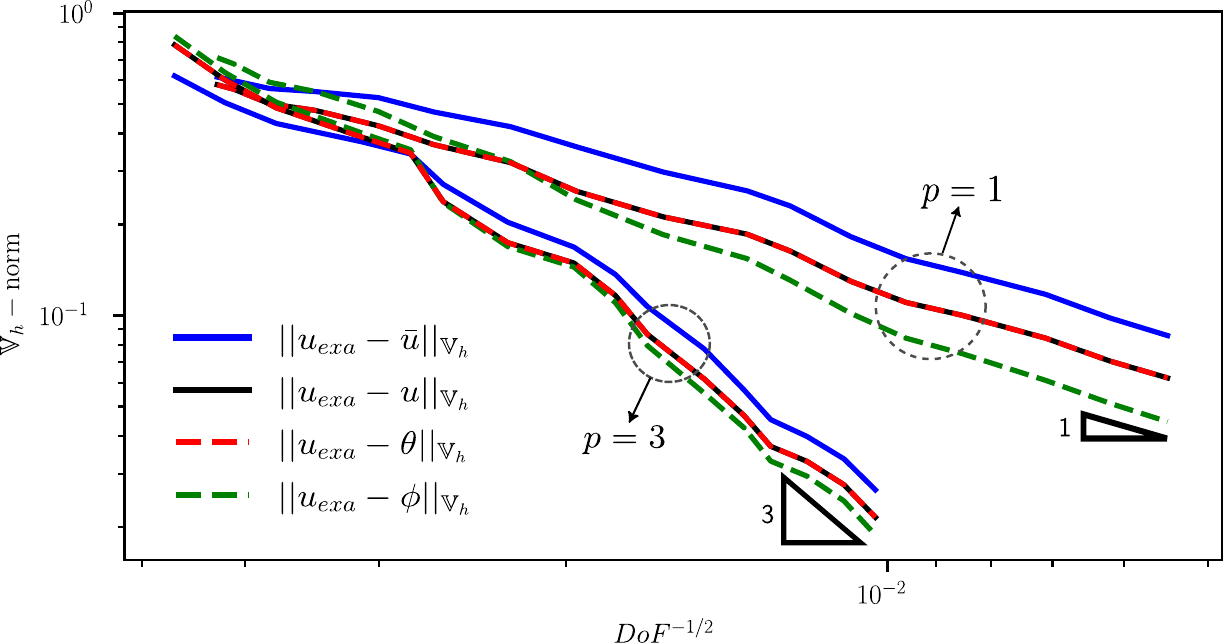}
			\caption{$\V_h$ norm}
		  \label{fg:EJS-vms-Vh-k1-3}
		\end{subfigure}
		\begin{subfigure}{0.495\textwidth}
			\centering
			\includegraphics[width=\textwidth]{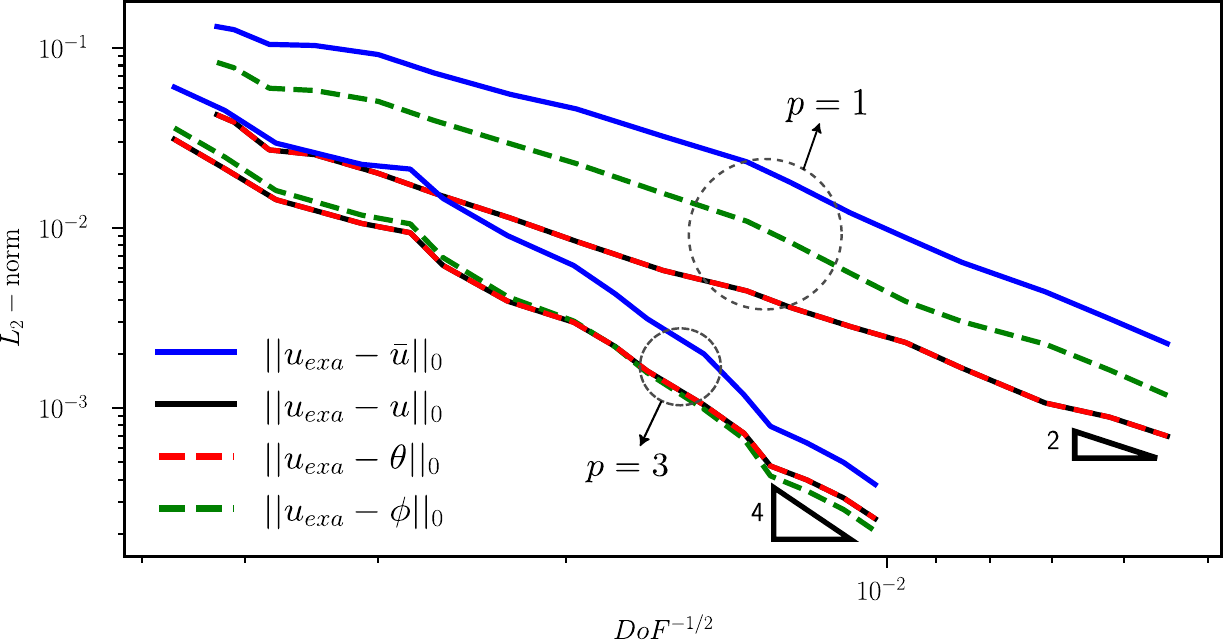}
			\caption{$L_2$ norm}
    		\label{fg:EJS-vms-L2-k1-3}
		\end{subfigure}
    	\caption{$\V_h-$ and $L_2-$norm convergence for 3D Eriksson-Johnson problem, $p=1,3$}
        \label{fg:ConvEJVh}
	\end{center}
\end{figure}

\begin{figure}[h!]
	\begin{center}
		\begin{subfigure}{0.495\textwidth}
			\centering
			\includegraphics[width=\textwidth]{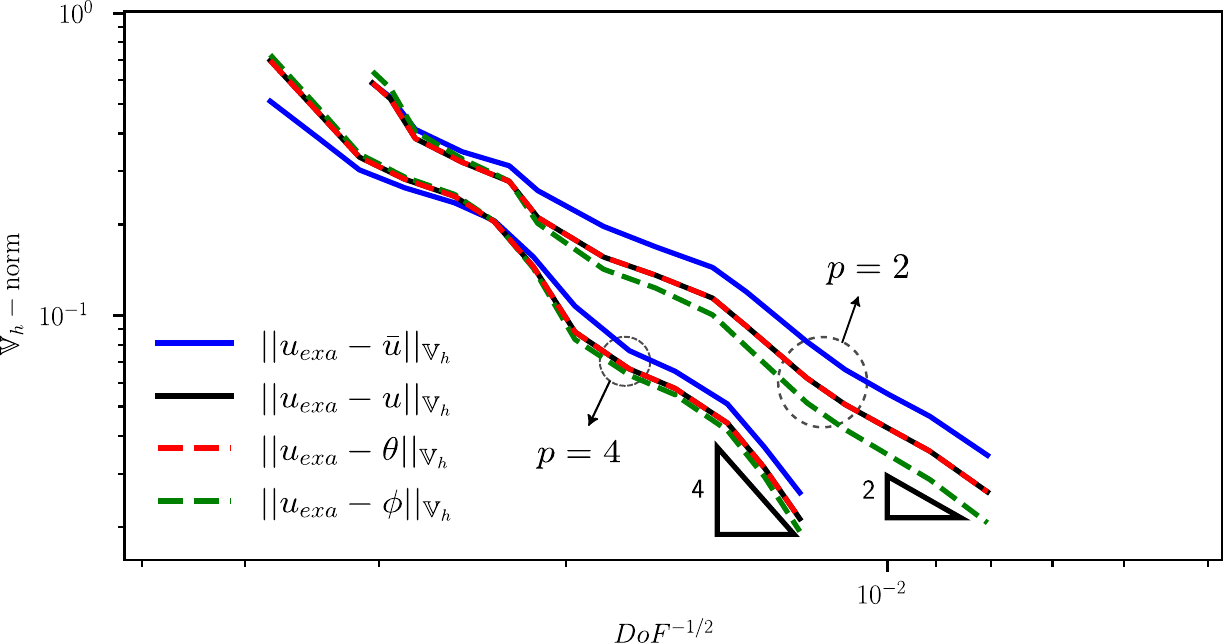}
			\caption{$\V_h$ norm}
            \label{fg:EJS-vms-Vh-k2-4}
        \end{subfigure}
		\begin{subfigure}{0.495\textwidth}
			\centering
			\includegraphics[width=\textwidth]{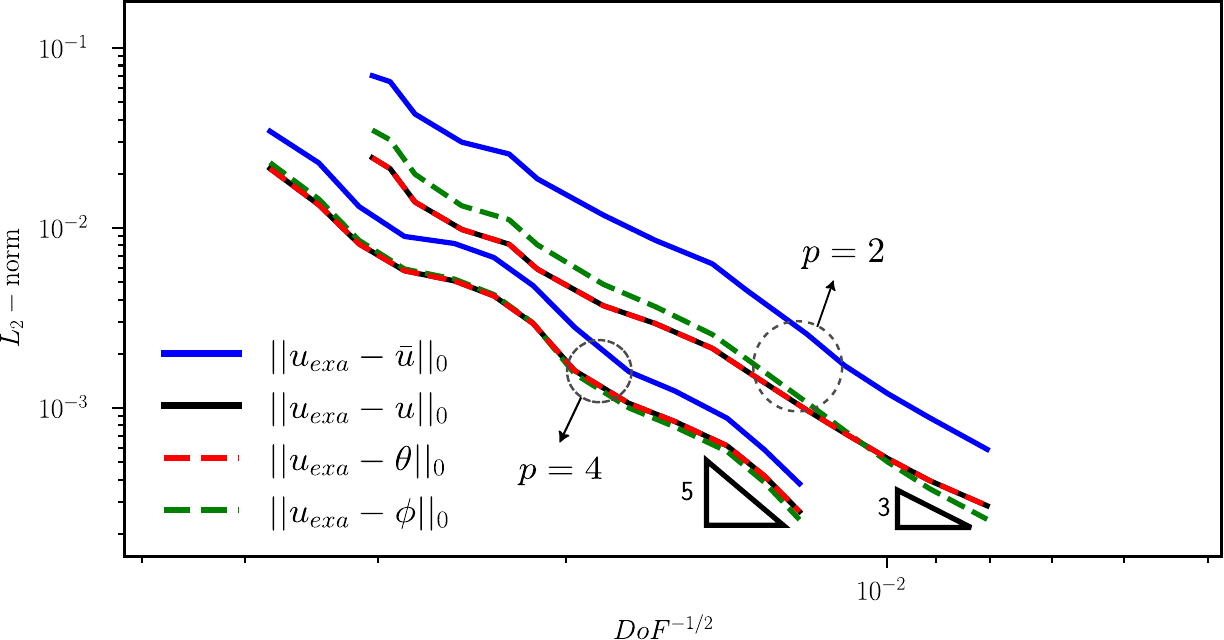}
			\caption{$L_2$ norm}
    		\label{fg:EJS-vms-L2-k2-4}
		\end{subfigure}
    	\caption{$\V_h-$ and $L_2-$norm convergence for 3D Eriksson-Johnson problem, $p=2,4$}
	    \label{fg:ConvEJL2}
	\end{center}
\end{figure}

\subsection{3D Eriksson-Johnson problem}
We use a 3D version of the classical Eriksson-Johnson problem described in~\cite{ Cier:2021, Chan:2013}. We solve the equation~\eqref{eq:DAR-HD} in the domain $\Omega = [0,1]^3$ with diffusion coefficient $\kappa =10^{-2}$, velocity field $\beta = (1,0,0)^T$ and a source term $f=0$. We impose Dirichlet boundary conditions from the analytical solution:
\begin{align*}
u_{exa}= \frac{\exp(r_1(x-1))-\exp(r_2(x-1))}{\exp(-r_1)-\exp(-r_2)} \sin(\pi y),
\end{align*}
with $r_{1,2} = {1 \pm \sqrt{1+4\kappa^2\pi^2}}/{2\kappa}$. 
Figure~\ref{fg:EJD} shows the coarse and fine scales for the problem for linear polynomials, $p=1$, and how the adaptive strategy captures the regions with sharp gradients, constructing a smooth solution. Figures~\ref{fg:ConvEJVh} and \ref{fg:ConvEJL2} show the optimal convergence plot in $\V_h$ and $L_2$ norms for the coarse- and full-scale approximations. As before, the adjoint multiscale reconstruction improves the $\V_h$ norm and the pre-asymptotic convergence, especially for lower-order polynomials.

\section{Extension to nonlinear conservation laws}
This section presents a numerical stability analysis for the nonlinear conservation law for our variational multiscale reconstructions. First, we formulate the discrete problem using the Lax-Friedrich flux for the nonlinear advective flux within the discontinuous Galerkin (dG) framework. Then, we describe the variational multiscale method for nonlinear problems~\cite{ Juanes:2005} and specialize it to the adaptive stabilized finite element method~\cite{ Calo:2020} to obtain a nonlinear fine-scale approximation. We present some numerical results for Burgers' equation to demonstrate the accuracy and efficiency of our approach.

\subsection{Discontinuous Galerkin discretization}
We consider the following nonlinear conservation law: 
\begin{equation}
\begin{aligned}
    \nabla \cdot \left( {\mathrm{f}(u)} - \kappa \nabla u  \right)&= f, &&\quad\text{in } \Omega, \\
		u &= u_D  , &&\quad\text{on } \Gamma_D,
	\label{eq:conservation-law}
\end{aligned}
\end{equation}
where ${\mathrm{f}(u)}$ represents a nonlinear convective flux; thus, the dG discrete problem reads:
\begin{equation}\label{eq:DGNL}
    \left\{
    \begin{array}{l}
	   \text{Find } u \in \V_h,  \text{ such that:} \smallskip\\
        \begin{aligned}
		  n_h(v;u)&= \ell_h(v), &&\quad \forall \, v \in \V_h,
        \end{aligned}
	\end{array}
	\right.
\end{equation}
where $n(v;u)$ represents the nonlinear form including a SWIP contribution in~\eqref{eq:swip} and the Lax-Friedrichs numerical flux $\Phi$ for the nonlinear convective flux defined as follows:
\begin{equation}
	\label{eq:eta_nl}
	n_h(v; u) = b_h(v,u)^{\text{swip}} -\sum_{K \in \mathfrak{T}}(\nabla_h v,\mathrm{f}(u))_K
	+  \sum_{F \in \Sk_h}([\![ v ]\!],\Phi )_F,
\end{equation}
with,
\begin{equation}
\begin{aligned}
\forall& F \in \Sk_h^0,  &&
\Phi :=  \frac{1}{2}(\mathrm{f}(u_1)\cdot \textbf{n}_F + \mathrm{f}(u_2)\cdot \textbf{n}_F + \eta_f (u_1 - u_2)),\\
\forall& F \in \Sk_h^D, &&
\Phi :=  \frac{1}{2}(\mathrm{f}(u)\cdot \textbf{n}_F +  \eta_f \,u) + \frac{1}{2}(\mathrm{f}(u_D)\cdot \textbf{n}_F - \eta_f \,u_D).
\end{aligned}
\end{equation}
$\eta_f$ is a local dissipation parameter and is chosen based on the maximum eigenvalue of the flux function Jacobian. In equation~\eqref{eq:Burgersa}, we set
\begin{equation}
\begin{aligned}
		\forall& F \in \Sk_h^0, && 
		\eta_f :=  \max_{w=u_1,u_2}|f'(w)\cdot \textbf{n}_F|, \\
		\forall& F \in \Sk_h^D, &&  
		\eta_f :=  |f'(u)\cdot \textbf{n}_F|.\\
\end{aligned}
	\label{eq:LDP}
\end{equation}
The right-hand side in~\eqref{eq:DGNL} for weakly imposed non-homogeneous Dirichlet boundary conditions reads:
\begin{align}
	\ell_h(v):
	= \sum_{K \in \mathfrak{T}}(v,f)
	&+ \sum_{F \in \Sk_h^{D}} \left( n_e\kappa(v, u_D)_F -  ( \kappa\nabla_h v \cdot \textbf{n}_F , u_D)_F \right) \nonumber\\ 
	&-\sum_{F \in \Sk_h^{D} \cap\Gamma} \tfrac{1}{2} \left(v , \mathrm{f}(u_D)\cdot \textbf{n}_F - \eta_f u_D \right) _F .
	\label{eq:lineear_form_NL}
\end{align}
We endow $\V_h$ with the norm:
\begin{equation}
	\begin{aligned}\label{eq:normVh-NL}
		\|w\|^2_{\V_h} :=  \|w\|^2_{\text{swip}} + \|w\|^2_{\text{conv}},
	\end{aligned}
\end{equation}
with $\|w\|^2_{\text{swip}}$ the SWIP norm contribution~\eqref{eq:swip-norm} and $\|w\|^2_{\text{conv}}$ a convective norm defined as follows: 
\begin{equation}
	\|w\|^2_{\text{conv}} := \displaystyle \|w\|^2_{0} + \displaystyle \sum_{ F \in \mathscr{S}_h}\|\llbracket w \rrbracket \|^2_{0,F} + \displaystyle \sum_{K \in \mathfrak{T}} h_K \| \, \nabla{w} \, \|^2_{0,K}.
	\label{eq:adv-normNL}
\end{equation}

\subsection{Residual minimization formulation}

Following the residual minimization formulation in Section~\ref{Section:Residual-minimization}, and the nonlinear approach for residual minimization problems~\cite{ Cier:2020}, we formulate the following nonlinear saddle-point problem to obtain the coarse-scale solution and an error estimate:
\begin{equation}
	\label{eq:saddle-point-NL}
	\left\{
	\begin{array}{l}
		\text{ Find } (\varepsilon, \bar u) \in \V_h \times \bar \V_h,  \text{ such that:} \smallskip \\
\begin{aligned}
			g(v,\varepsilon) + n_h(v;\bar u) &= l_h(v), &&\quad\forall v \in  \V_h, \smallskip\\
			n'_h(\varepsilon,\bar w ; \bar u) &= 0,&& \quad\forall \bar w \in \bar \V_h. \smallskip
\end{aligned}
	\end{array}
	\right.
\end{equation}
Here, $n_h'(v,u;\delta u)$ is the linearized form of the nonlinear operator~\eqref{eq:eta_nl}, evaluated at $\bar u\in\V_h$, and defined through the discrete G\^ateux derivative in the direction $\delta\bar u\in\V_h$ as:
\begin{equation}
n_h\sp{\prime}(v; \delta\bar u, \bar u):=\frac{d}{d\epsilon}n_h(v ;\bar u+\epsilon \delta\bar u)\big|_{\epsilon=0.}
\end{equation}
We use the Newton-Raphson method to solve~\eqref{eq:saddle-point-NL} so that at each step of the nonlinear iteration, we solve the following linear problem:

\begin{equation}
	\label{eq:saddle-point2}
	\left\{
	\begin{array}{l}
		\text{Given } (\varepsilon_i, \bar u_i), \text{ find } (\delta \varepsilon, \delta \bar u) \in \V_h \times \bar \V_h,  \text{ such that:} \smallskip \\
		\begin{array}{lcll}
			g(v,\delta \varepsilon) + n'_h(v,\delta \bar u ; \bar u_i ) &=& l_h(v) - g(v, \varepsilon_i) - n_h(v,\bar u_i ) & \quad \forall v \in  \V_h, \smallskip\\
			n'_h(\delta \varepsilon,\bar w ; \bar u_i) &=& -n'_h( \varepsilon_i,\bar w ; \bar u_i) & \quad \forall \bar w \in \bar \V_h, \smallskip\\ 
			
		\end{array}
	\end{array}
	\right.
\end{equation}
where we update $u_i$ and $\varepsilon_i$ at every iteration $i$ such that:
\begin{align}
\label{eq:increment}
 \bar u_{i+1} &=  \bar u_i + k \delta \bar u , && \varepsilon_{i+1} =  \varepsilon_i + k \delta \varepsilon
\end{align}
and assume convergence when $\| \bar{u}_{i+1} - \bar{u}_{i} \|_{\V_h} < 10^{-6}$.
We use a damped Newton algorithm to solve the corresponding nonlinear saddle point problem. We denote the relaxation parameter as $k$ in~\eqref{eq:increment}, following~\cite{ Bank:1981} (see algorithmic details in~\cite{ Cier:2020}).
Regarding error estimation, we adopt a similar approach to that used in the linear problem. Here, $\varepsilon$ in equation~\eqref{eq:saddle-point-NL} represents the residual error of the nonlinear operator, i.e., $g(v,\varepsilon) = l_h(v) - n_h(v;\bar u)$. However, we use the updated residual error ($\varepsilon_{i+1}$) to guide adaptivity after the solution of~\eqref{eq:saddle-point2} converges.

\subsection{Nonlinear variational multiscale method}

Following the previous section, a natural approach to solve ~\eqref{eq:DGNL} uses the resulting linearized equation and updates the solution with a correction term at every iteration as:

\begin{equation}
\label{eq:increment-2}
u_{i+1} = u_i + \delta u.
\end{equation}
In a multiscale context, both $u_i$ and $\delta u$ in~\eqref{eq:increment-2} are decomposed  into fine-scale ($\tilde u_i$, $\delta \tilde u$) and coarse-scale ($\bar u_i$, $\delta \bar u$) components. To circumvent the need for decomposing $u_i$ at each iteration, we use a decoupled scale system by employing the subsequent approximation, as suggested by ~\cite{Juanes:2005}:

$$u_i \approx \bar u_i,$$
which assumes the following at each iteration,
\begin{equation}
\label{eq-uaprox}
u_{i+1} \approx \bar u_{i+1} + \delta \tilde u.
\end{equation}
We simplify the notation in ~\eqref{eq-uaprox} by dropping the sub-index $i$:
\begin{equation}
\label{eq-uaprox2}
u \approx \bar u + \delta \tilde u.
\end{equation}
From this, we define a multiscale formulation through defining a nonlinear direct sum decomposition for both the trial and test function spaces:
\begin{align}
\label{eq:full-scale-NL}
u &= \bar u + \delta \tilde u && \forall u \in \V_h,\\
v &= \hat v + v' &&\forall v \in \V_h,
\end{align}
where $\bar u\in\bar\V_h$, $\delta \tilde u \in \tilde \V_h$, $\hat v\in\hat\V_h$, and $v'\in\V_h'$.  Using~\eqref{eq:full-scale-NL}  in the nonlinear problem~\eqref{eq:DGNL}, and a first-order Taylor expansion about the coarse-scale solution $\bar u$, the nonlinear form becomes:
\begin{equation}
\label{eq:linear-expansion}
n_h(v;u) = n_h(v;\bar u + \delta \tilde u) \approx n_h(v;\bar u) + n'_h(v;\delta \tilde u , \bar u)   \quad  \, v \in  \V_h.
\end{equation}
	Drawing from the first equation in the nonlinear saddle-point problem in ~\eqref{eq:saddle-point-NL} and employing ~\eqref{eq:DGNL} and~\eqref{eq:linear-expansion}, we obtain the following identity: 
	
	\begin{equation}\label{eq:err-estimative}
		\left\{
		\begin{array}{l}
			\text{Find } \delta \tilde u  \in \V_h , \text{  such that:  }   \smallskip \\
			\begin{aligned}
				n'_h(v,\delta \tilde {u};\bar{u}) &= g(v,\varepsilon) &&\forall v \in \V_h.
				\end{aligned}
			\end{array}
		\right.
	\end{equation}
Here, we use $\varepsilon$ and $\bar u$ from ~\eqref{eq:saddle-point-NL} after the system converges. Similar to the linear case, in the subsequent sections, we represent the reconstructed full-scale solution (Equation~\eqref{eq:full-scale-NL}) as ($u$) and the dG solution in ~\eqref{eq:DGNL} as ($\theta$). Unlike the linear problem, $u$ is not equivalent to, but rather an approximation of, $\theta$ (i.e., $u \approx \theta$).\\
The decoupled multiscale formulation substantially reduces computational costs by eliminating the need for explicit decomposition of $u_i$ into coarse- and fine-scale components during each iteration. Instead, the decomposition is only performed once at each iteration level. In the next section, we demonstrate that our approach and the adaptive strategy yield an accurate and stable approximation of the coarse-scale solution and error indicator. This approximation enables the recovery of the full-scale solution following refinement.

\subsection{Adjoint multiscale reconstruction}

As for the linear case, we introduce an adjoint multiscale reconstruction $\phi$ defined as:
$$\phi = \bar u + \delta \check u,$$
where $\delta \check u$ is fine-scale adjoint reconstruction, such that: 
\begin{equation}\label{eq:res-dgprimeNL}
	\left\{
	\begin{array}{l}
		\text{Find } \delta \check u  \in \V_h , \text{  such that:  }   \smallskip \\
		\begin{aligned}
			n'_h(v,\delta \check {u};\bar{u}) = g(v,\varepsilon') + 
			n'_h(\varepsilon',v;\bar{u}) &&\forall v \in \V_h.
		\end{aligned}
	\end{array}
	\right.
\end{equation}
The following example shows the improved accuracy of the adjoint reconstruction ($\phi$) for the asymptotic regime in the energy norm (i.e., $||u_{exa} - \phi ||_{\V_h} \lesssim ||u_{exa} - u ||_{\V_h} $).


\begin{figure}[h!]
	\begin{center}
		\begin{subfigure}{0.49\textwidth}
			\centering
			\includegraphics[width=0.75\textwidth]{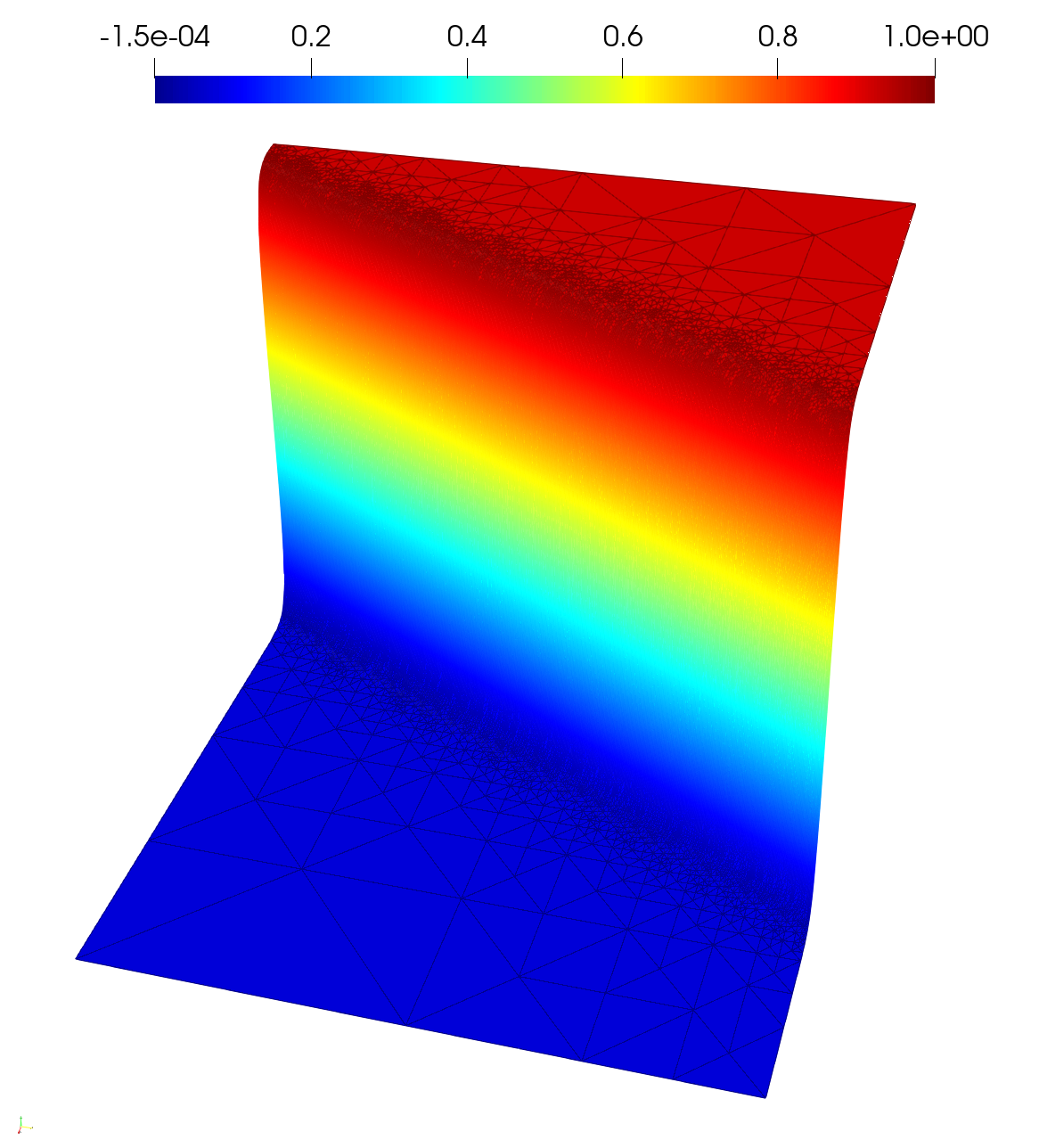}
            \caption{Coarse-scale solution}
			\label{fg:advection-coarse}
		\end{subfigure} 
		\begin{subfigure}{0.49\textwidth}
			\centering
			\includegraphics[width=\textwidth]{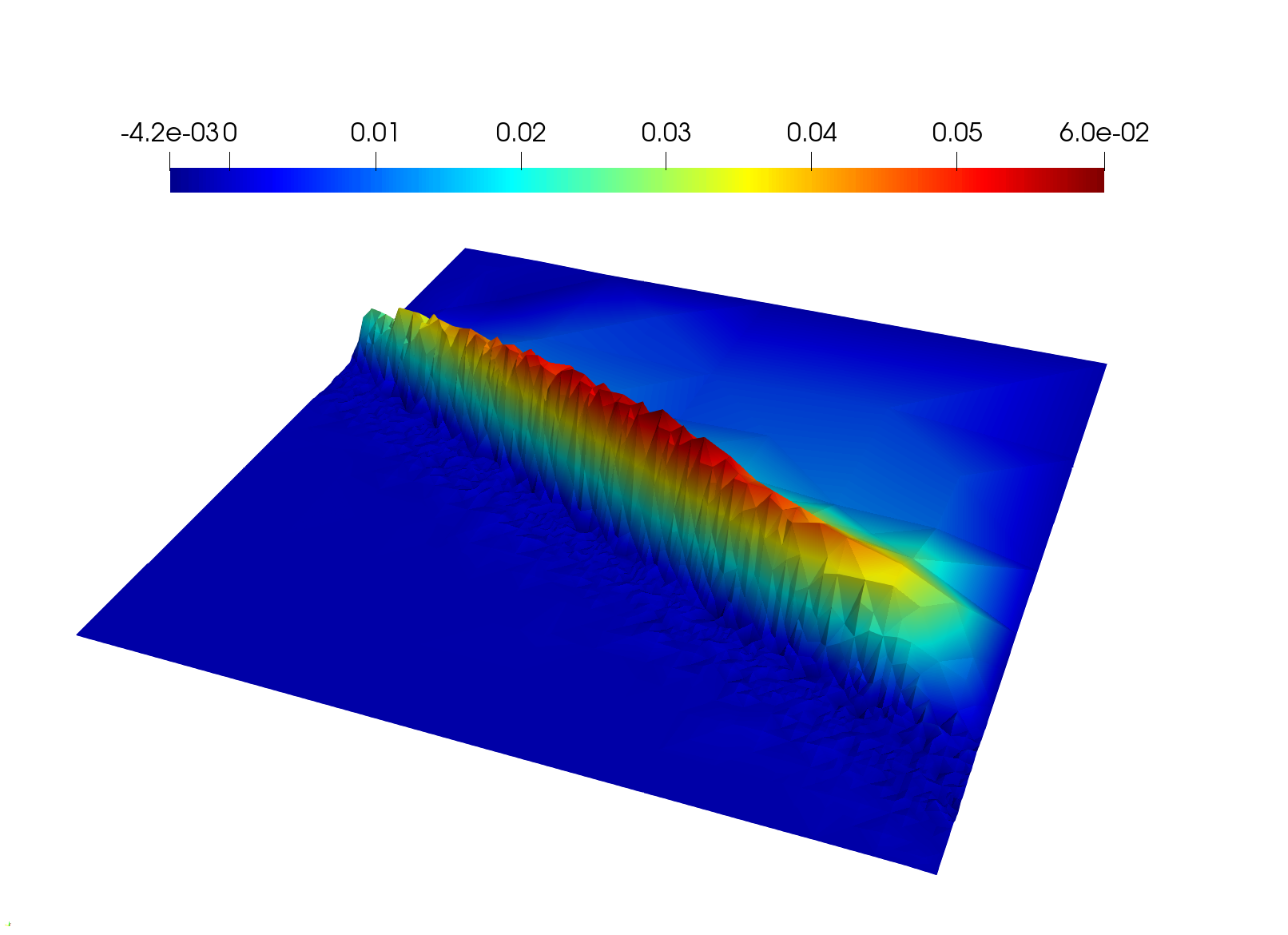}
            \caption{Fine-scale solution}
			\label{fg:advection-fine}
		\end{subfigure} 
		\caption{Solution for Burgers' equation for $\kappa=10^{-3}$ \& $p=1$}
		\label{fg:burges-iso-fig}
	\end{center}
\end{figure}

\begin{figure}[h!]
    \centering
	\includegraphics[width=1\textwidth]{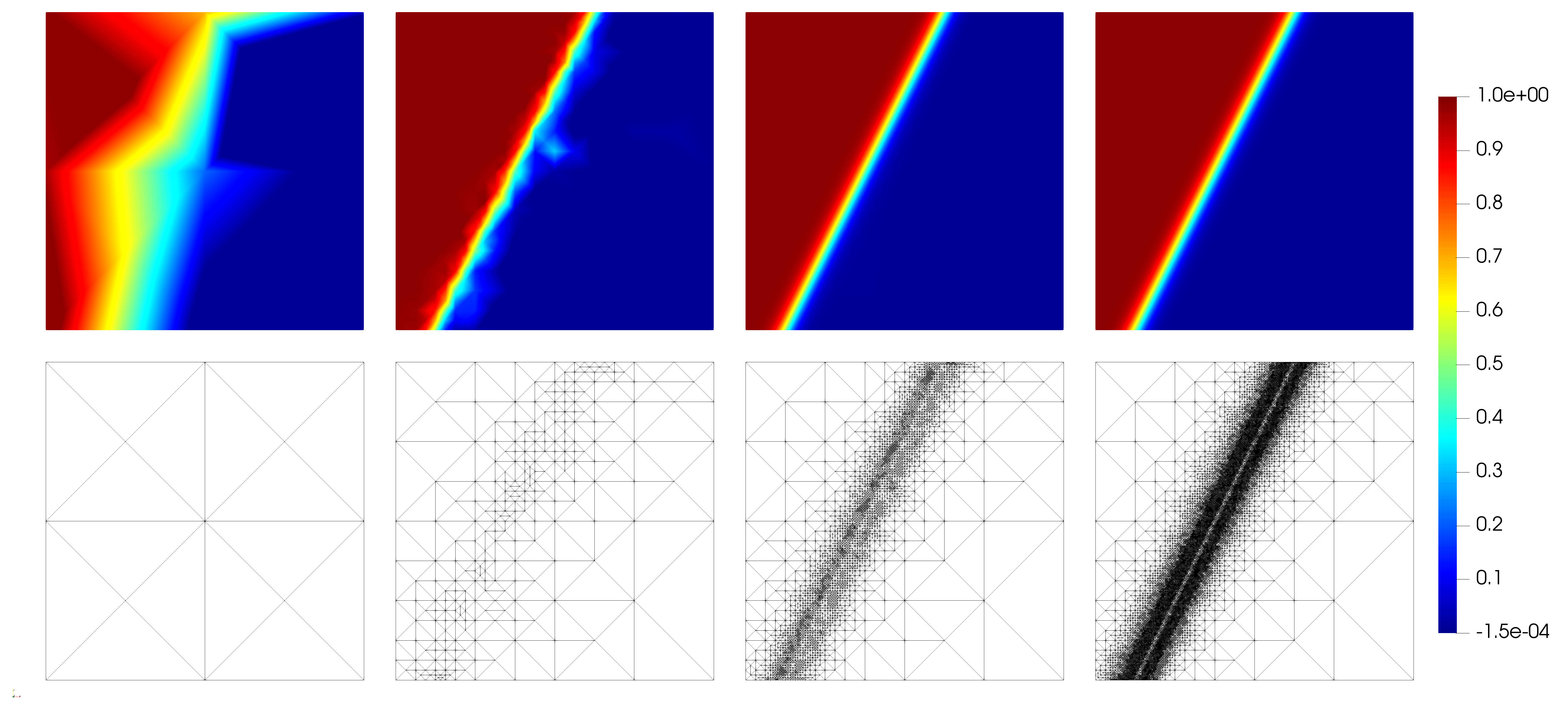} \\
 	\scriptsize {\hspace{-0.5cm}  Level= 0 \hspace{2.2cm} Level= 10 \hspace{2.2cm} Level= 20 \hspace{2.1cm} Level= 25 \hspace{0.2cm}} 

	\caption{Solution for the isotropic Burgers equation for different refinement levels at $\kappa=10^{-3}$}
	\label{fg:iso-burgers-sequence}
\end{figure}

\begin{figure}[h!]
\centering
		\begin{subfigure}{0.49\textwidth}
			\includegraphics[width=\textwidth]{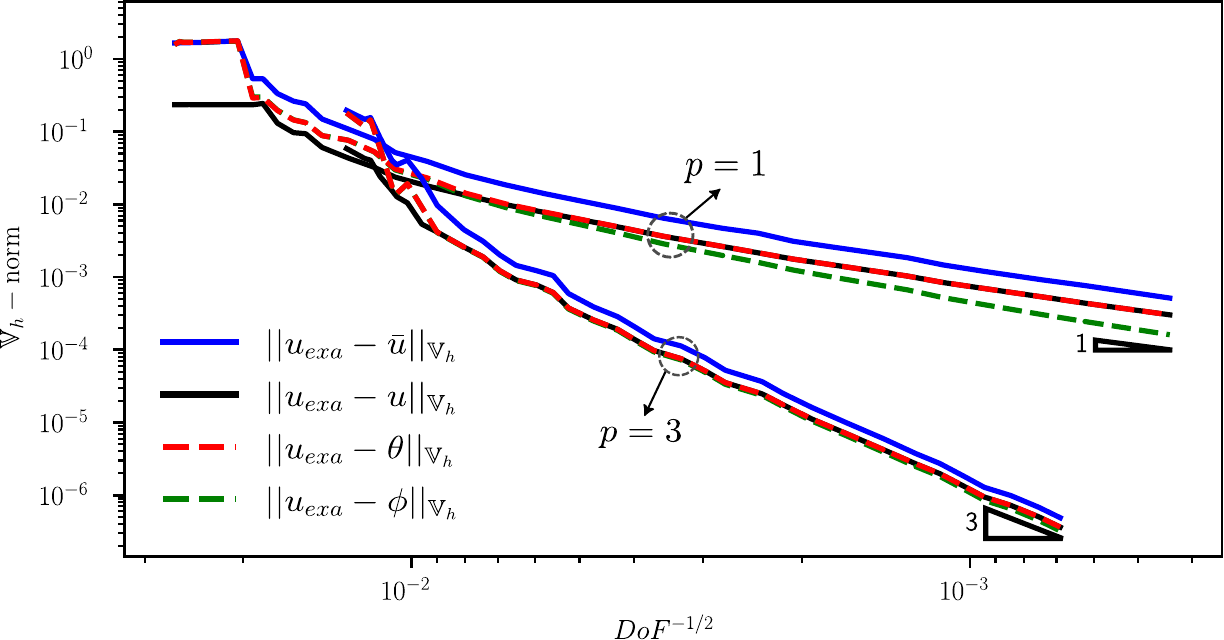}
			\caption{$\V_h$ norm}
			\label{fg:burgers-k1-3-Vh}
		\end{subfigure} 
		\begin{subfigure}{0.49\textwidth}
			\includegraphics[width=\textwidth]{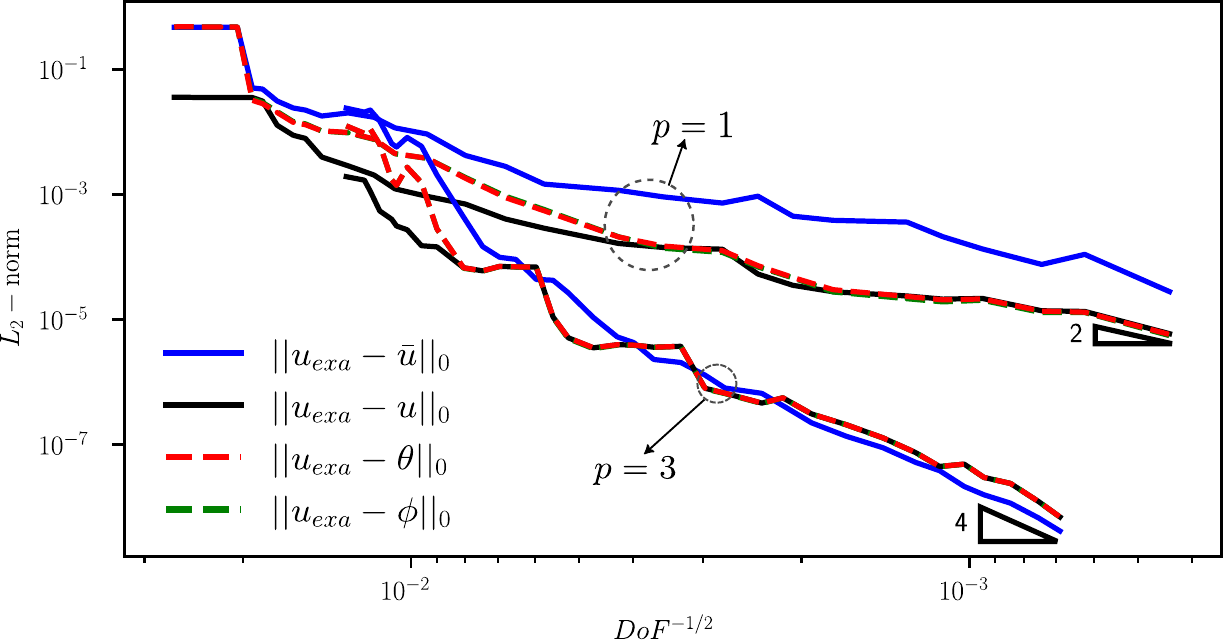}
			\caption{$L_2$ norm}
			\label{fg:burgers-k1-3-L2}
		\end{subfigure}
		\caption{Convergence plots for Burgers' equation in the $L_2$ and $\V_h$ norms, $p=1,3$ }
		\label{fg:Burgers-convergence-k1-3}
\end{figure}

\begin{figure}[h!]
\centering
		\begin{subfigure}{0.49\textwidth}
			\includegraphics[width=\textwidth]{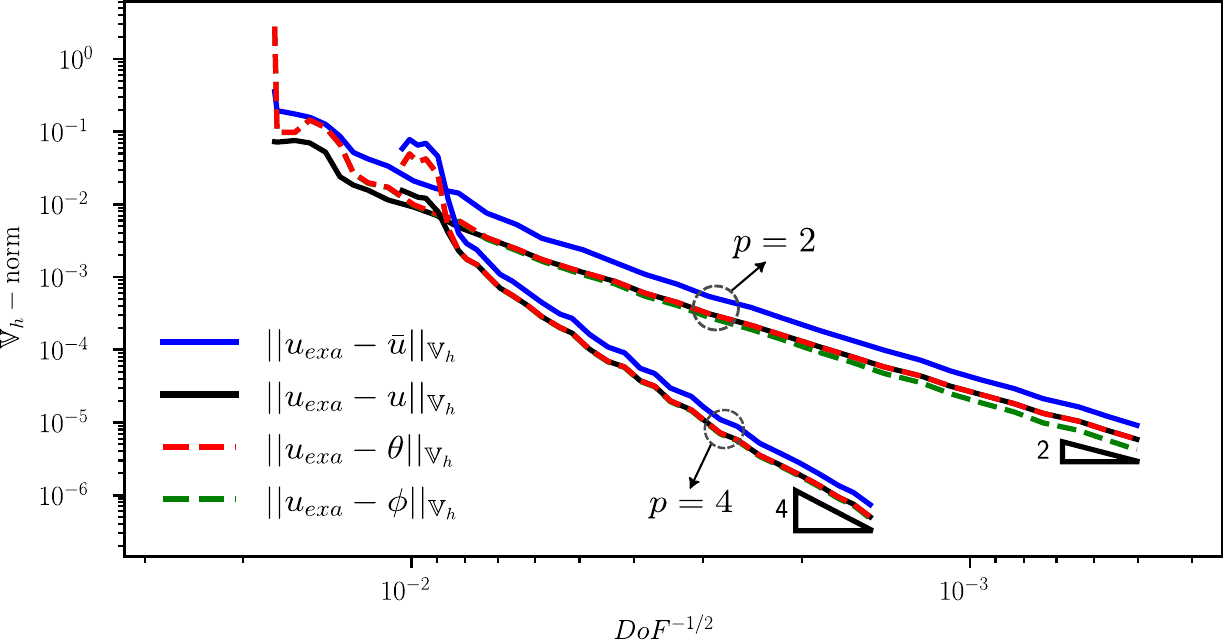}
			\caption{$\V_h$ norm}
			\label{fg:burgers-k2-4-Vh}
		\end{subfigure} 
		\begin{subfigure}{0.49\textwidth}
			\includegraphics[width=\textwidth]{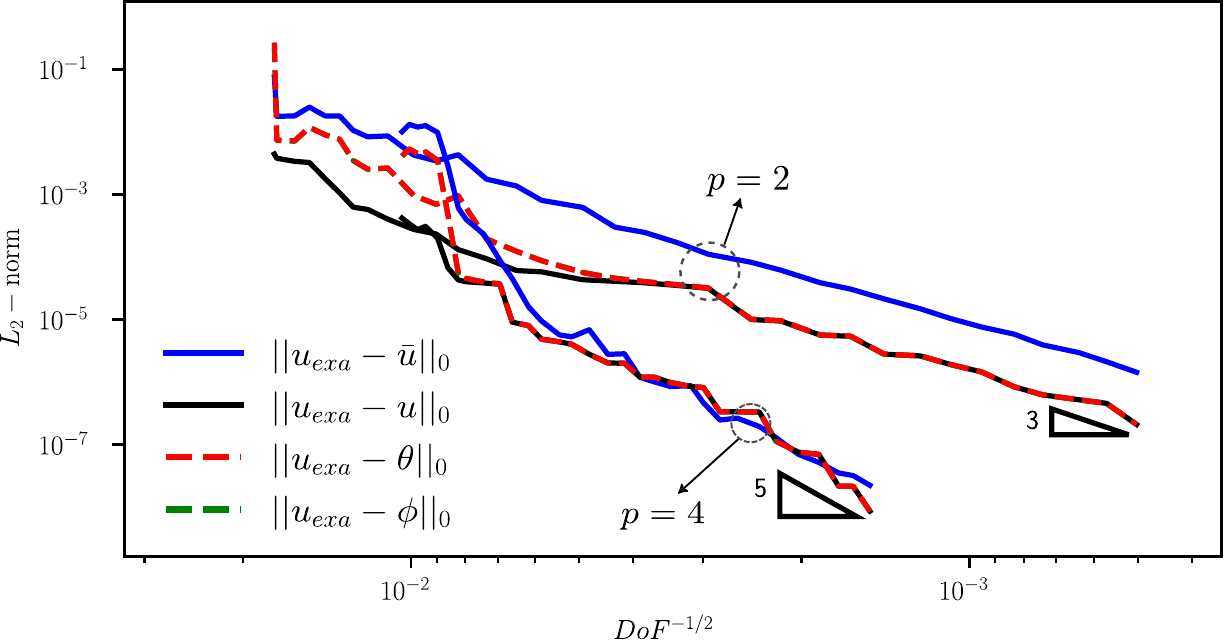}
			\caption{$L_2$ norm}
			\label{fg:burgers-k2-4-L2}
		\end{subfigure}
		\caption{Convergence plots for Burgers' equation in the $L_2$ and $\V_h$ norms, $p=2,4$  }
		\label{fg:Burgers-convergence-k2-4}
\end{figure}

\subsection{Burgers' equation}

This section assesses the performance of the proposed variational multiscale reconstructions applied to a nonlinear problem. We focus on the steady Burgers' equation, a widely used partial differential equation for modelling physical phenomena, including fluid dynamics, shock waves, and traffic flow. Subsequent sections will showcase results derived from this equation's isotropic and anisotropic variations.

\subsubsection{Isotropic Burgers' equation}
In the first example, we solve the isotropic case where $\mathrm{f}(u):= {\frac{\textbf{b} u^2}{2}}$  with $\textbf{b}=[1,1]^T$. The problem reads:
\begin{equation}
	\begin{array}{rlll}
		\nabla \cdot \left(  {\frac{\textbf{b} u^2}{2}} \right)  - \kappa \Delta u &=& f, \quad &\text{in } \Omega, \\
		u &=& u_D  , \quad &\text{on } \Gamma_D,
	\end{array}
	\label{eq:Burgersa}
\end{equation}
with $\Omega=[0,1]^2$, $\kappa=10^{-3}$ and a initial condition $u_0 = 0.5$ which rises to a inner discontinuity. We impose the source term $f$ and Dirichlet boundary condition ($u_D$) from the exact solution:
$$ u_{exa} = \frac{1}{2} \left (1-\tanh \left( \frac{2x-y-0.25}{\sqrt{5\kappa}} \right) \right). $$   
We use $f'(w) = \textbf{b}w$ to impose the local dissipation parameter in~\eqref{eq:LDP}.

\begin{figure}[h!]
    \centering
	\includegraphics[width=0.8\textwidth]{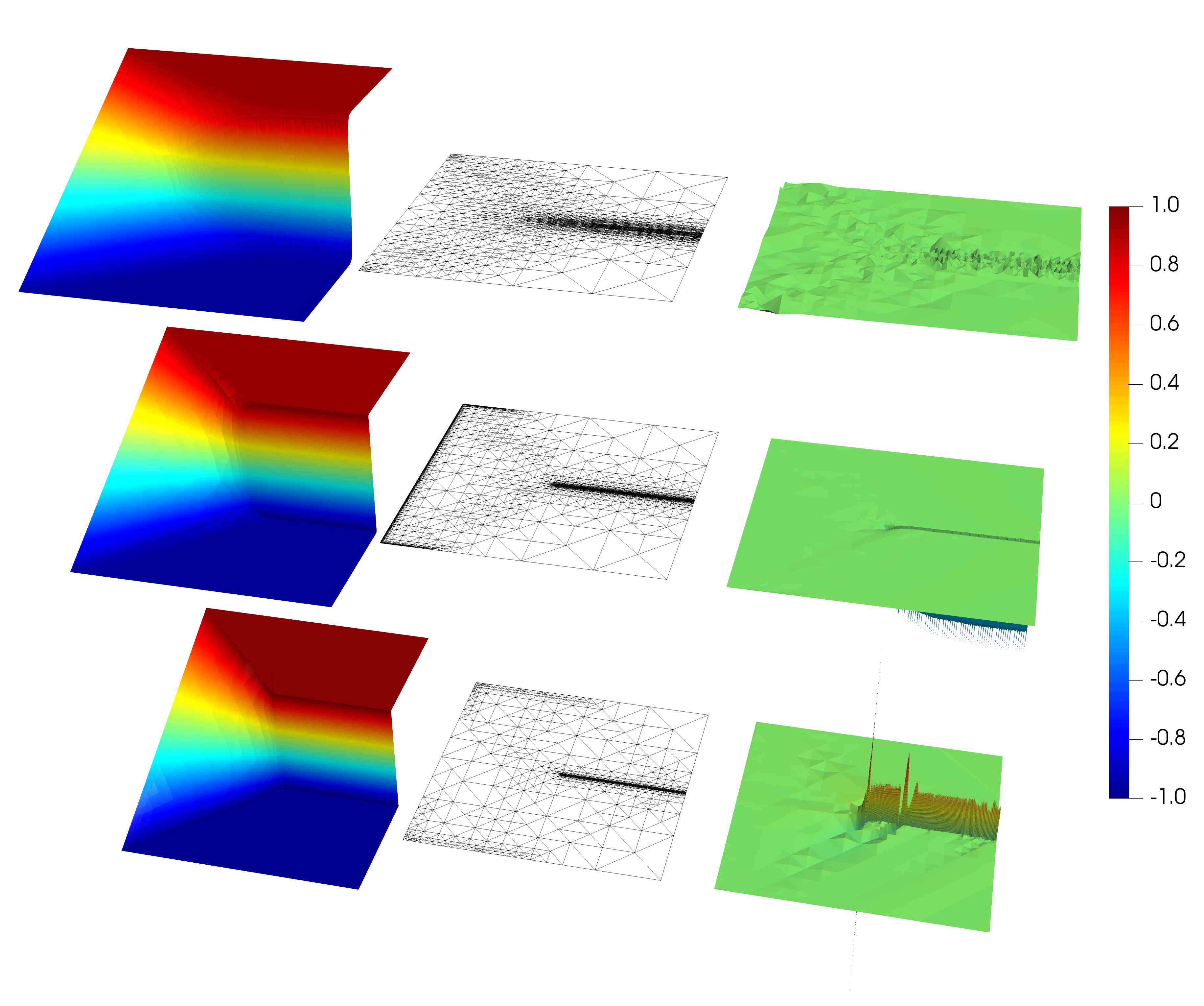}
 \\
 \vspace{-0.5cm}
 	\scriptsize {
  \hspace{-0.5cm} (a) Coarse-scale solution 
  \hspace{1.2cm} (b) Mesh \hspace{1.2cm} (c) Fine-scale solution  \hspace{0.2cm}} 
	\caption{Coarse-scale solution and scale approximations for different diffusivities. (top: $\kappa=10^{-2}$, middle: $\kappa=10^{-3}$ ,bottom: $\kappa=10^{-4}$)}
 
	\label{fg:ani-burgers}
\end{figure}

\begin{figure}[h!]
    \centering
	\includegraphics[width=.8\textwidth]{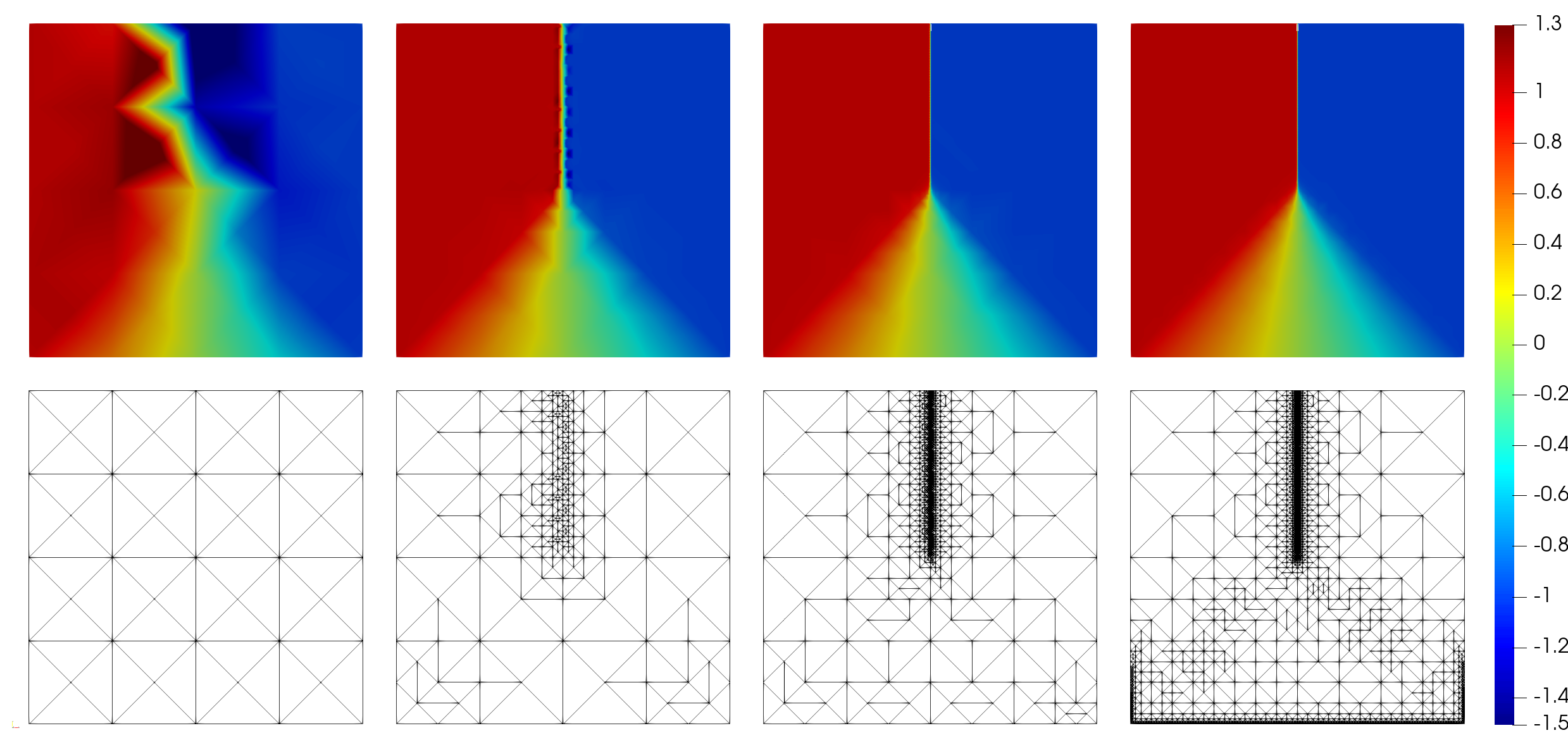} \\
 	\scriptsize {Level= 0 \hspace{1.5cm} Level= 5 \hspace{1.5cm} Level= 10 \hspace{1.5cm} Level= 20 \hspace{0.2cm}} 

	\caption{Solution for the Burgers equation for different refinement levels at $\kappa=10^{-2}$}
	\label{fg:ani-burgers-sequence}
\end{figure}
\noindent Figure~\ref{fg:burges-iso-fig} shows solution plots for $p=1$, illustrating the smooth approach of the continuous solution and its correction for the fine-scale along the discontinuity. Figure \ref{fg:iso-burgers-sequence} displays a sequence of four refinement levels to demonstrate the effectiveness of the refinement strategy in capturing the sharp inner layer, irrespective of the initial mesh.
Figures~\ref{fg:Burgers-convergence-k1-3} and~\ref{fg:Burgers-convergence-k2-4} depict optimal convergence rates for $p=1,2,3,4$ in the $L_2$ and $\V_h$ norms. Moreover, we show our method recovers the full-scale solution (i.e., $u \approx \theta$) after refinement, regardless of the polynomial order. Furthermore, as in the linear case, the adjoint multiscale reconstruction ($\phi$) enhances the approximation quality in the $\V_h$ norm.

\subsubsection{Single-component Burgers' equation}

We introduce a second scenario to test the numerical performance in the presence of a sharp shock layer. We select a problem from~\cite{ Moro:2012} to solve Burgers' equation in a single component, such that  $\mathrm{f}(u):= \left[{\frac{u^2}{2}},u \right]^T$.
This problem represents a challenge, especially for convection-dominated problems, due to the shock presence at $x=0$.
The problem reads:
\begin{equation}
\frac{1}{2} \frac{\partial u^2}{\partial x} + \frac{\partial u}{\partial y} =   \kappa \left(\frac{\partial^2 u}{\partial x^2} + \frac{\partial^2 u}{\partial y^2}  \right), \quad \quad \text{in } \Omega, 
\label{eq:Burgersa2}
\end{equation}
with $\Omega=[0,1]^2$ and initial guess $u_0 = 1-2x$. We impose Dirichlet boundary conditions $u_D = u_0$ at $y=0$, $x=0$ and $x=1$ and zero Neumann boundary conditions at $y = 1$. The local dissipation parameter in~\eqref{eq:LDP} uses $f'(w) = [u,1]$. We used a polynomial degree of $p=3$ and a uniform element size of $4\times4$ for the initial mesh. Figure~\ref{fg:ani-burgers} illustrates the coarse and fine solutions for $\kappa=10^{-2}$, $\kappa=10^{-3}$, and $\kappa=10^{-4}$. Additionally, Figure \ref{fg:ani-burgers-sequence} demonstrates how the adaptive strategy effectively reduces oscillations along the shock for coarse meshes, regardless of the initial mesh~\cite{ Moro:2012}. 

\section{Conclusions}\label{sc:concl} 
We introduce an adaptive stabilized finite element method for convective-dominated diffusion problems using variational multiscale fine-scale reconstructions. The method constructs a coarse-scale approximation by minimizing the residual in dual discontinuous Galerkin norm and employs a robust error estimation technique to obtain a fine-scale solution and guide adaptivity. We show the effectiveness and reliability of our method through various numerical experiments involving different types of linear problems, such as highly heterogeneous, strong anisotropic diffusion tensors, and convection-dominated diffusion. Lastly, we successfully extend the method to solve the nonlinear conservation law. We use Burgers' equation as an example where we obtain stable solutions and optimal convergence rates.

\section{Acknowledgments}
This work is supported by the close collaboration between Curtin University and CSIRO. under the CSIRO DEI FSP postgraduate top-up
scholarship (Grant no. 50068868). J.G. gratefully acknowledges Roberto Rocca Education Program for its support.

\bibliography{bibfile}  

\end{document}